\documentclass[referee]{aa}

\usepackage{graphicx}
\usepackage{amsmath,epsfig,rotating,amssymb}

\begin{document}

\def \rot{{\rm {\bf rot} }}
\def \grad{{\rm {\bf grad} }}
\def \div{{\rm div}}
\def \cha{\widehat}
\def \pr{{\it permanent}  regime }


\author{Audit E. \inst{1} and Hennebelle P. \inst{2} }

\institute{  Service d'Astrophysique, CEA/DSM/DAPNIA/SAp, C. E. Saclay,
  \newline F-91191 Gif-sur-Yvette Cedex
\and Laboratoire de radioastronomie millim{\'e}trique, UMR 8112 du
CNRS, 
\newline {\'E}cole normale sup{\'e}rieure et Observatoire de Paris,
 24 rue Lhomond,\newline 75231 Paris cedex 05,
France }

\offprints{E. Audit, P. Hennebelle  \\
{\it e-mail:} edouard.audit@cea.fr, patrick.hennebelle@ens.fr}   

\title{Thermal condensation in a turbulent atomic hydrogen flow}
\titlerunning{Thermal condensation in atomic hydrogen}

\abstract{ We present a numerical  and analytical study of the thermal
fragmentation of a turbulent  flow of interstellar hydrogen.  We first
present  the different  dynamical  processes and  the  large range  of
spatial (and  temporal) scales that need to  be adequately represented
in numerical simulations.  Next, we present bidimensionnal simulations
of turbulent converging flows which induce  the dynamical condensation
of the  warm neutral phase into  the cold phase.  We  then analyse the
cold structures and the fraction  of unstable gas in each simulation,
paying  particular  attention  to  the  influence  of  the  degree  of
turbulence.  When the  flow is very turbulent a  large fraction of the
gas remains in the thermally unstable domain.  This unstable gas forms
a  filamentary  network.   We  show  that the  fraction  of  thermally
unstable gas  is strongly correlated  with the level of  turbulence of
the  flow.  We  then develop  a semi-analytical  model to  explain the
origin of this  unstable gas.  This simple model  is able to reproduce
quantitatively  the   fraction  of   unstable  gas  observed   in  the
simulations and  its correlation  with turbulence. Finally,  we stress
the fact that even when the flow is very turbulent and in spite of the
fact that a large fraction of the gas is maintained dynamically in the
thermally unstable  domain, the classical picture of  a 2-phase medium
with stiff thermal fronts and  local pressure equilibrium turns out to
be   still  relevant  in   the  vicinity   of  the   cold  structures.
 }

\maketitle

\section{Introduction}
The  dynamics  of the  interstellar  gas  is  of great  importance  to
understand the formation of structures such as molecular clouds and it
is therefore determinant in the star formation process.

In this respect, the neutral atomic  hydrogen (HI) is crucial since it
represents more than 50$\%$ of the  ISM and since the molecular clouds
form through  the  condensation of HI  gas.   Because of the numerical
stiffness of the    problem, previous numerical models  attempting  to
simulate the ISM at a scale of about 1kpc and  to form molecular clouds
self-consistently  have  not  considered heating and  cooling functions
that  allow   thermal bistability  between  100   and   8000  K  (e.g
V\'azquez-Semadeni  et    al.    1995, 1996,   Korpi    et al.   1999,
Ballesteros-Paredes et al.  1999) or have a numerical resolution which
is not appropriate  to  adequately describe  this physics down  to the
scale of  the CNM  structures (Gazol  et al.  2001).  Therefore  it is
currently  unclear and indeed  almost unexplored  to  what extent  the
physics of HI may  or  may not have  a  significant influence  on  the
formation and the evolution of molecular clouds.

From the  theoretical point   of  view (Field  et  al. 1969,  McKee \&
Ostriker 1977, Wolfire et al. 1995)  as well as from the observational
one (Low et al. 1984, Boulanger  \& P\'erault 1988, Kulkarni \& Heiles
1987, Troland \& Heiles 2003), it is now well established that HI is a
thermally bistable  medium   which at thermal  equilibrium   and for a
thermal pressure close  to about (in the  vicinity of the sun)  4000 K
cm$^{-3}$, can   be in two different   thermodynamical states, namely a
warm and diffuse phase (WNM) and a cold and dense one (CNM) roughly in
pressure equilibrium.

The linear stability analysis  (Field 1965), the  quasi-static thermal
front propagation  (Zel'dovich  \&  Pikel'ner 1969, Penston   \& Brown
1970)  and  more generally   the non-linear  development of  a  single
structure (see Meerson 1996 for a  review and S\'anchez-Salcedo et al.
2002 for a recent study) have been under investigation for a long time
and are reasonably well understood.  However, it is only recently that
the behaviour of  a thermally bistable  flow in the fully dynamical or
turbulent regime has been investigated.

\subsection{Previous work}
In  the    context  of solar   chromosphere    and corona, Dahlburg et
al. (1987)  and Karpen et  al. (1988) perform  2D simulations of a gas
that  may undergo thermal   instability.  They have considered  random
velocity perturbations   in  an initially thermally   unstable gas and
study the development   of  cold  structures.  They  note  that  large
amplitude  velocity   perturbations can    prevent  the formation   of
condensed structures.

Kovalenko \& Shchekinov (1999) and Hennebelle \& P\'erault (1999) have
simultaneously investigated the possibility that a converging flow may
dynamically trigger the thermal  transition  from the  WNM  (thermally
stable) phase into the CNM phase. They perform 1D simulations and show
that if the perturbation lasts long enough (more  than a cooling time)
and is strong enough (velocity must be comparable to the sound speed),
the  thermal  transition is   possible,  i.e part   of the  WNM  phase
condenses  in CNM  leaving  a cold   structures   embedded in  a  warm
surrounding medium.  Their underlying idea is that an external forcing
like   bubble expansion  or any   phenomena   generating systematic or
turbulent  motions  may promote the  formation  of cold structures.  A
similar picture has been investigated by Koyama \& Inutsuka (2000) who
simulate  a shock propagating   in   HI and include H$_2$   formation.
Hennebelle  \& P\'erault (2000) have  also  considered the  case of  a
magnetic   field,  important in   the context  of   the  ISM, which is
initially oblique with  respect to the  velocity  field. They  show that
whatever the  value of the magnetic  field the thermal condensation is
still possible provided the initial  angle between ${\bf B}$ and ${\bf
V}$  is  small   enough.   The smallest angle  below     which thermal
condensation is always possible, is about  20$^\circ$. Due to magnetic
tension the  condensation is more  difficult for intermediate magnetic
intensity  than for a stronger field.

This paradigm  has been  further investigated  by Koyama  \&
Inutsuka (2002) who  simulate  in 2D a  shock propagating  in HI. They
show that several structures of CNM form close  to the shock interface
and find that the velocity dispersion of  the cold structures is about
5-6 km/s, i.e a fraction of the sound speed of the WNM.

Gazol et al.  (2001) have performed 2D  simulations of HI at a scale  of
about 1 kpc and a numerical resolution of  about 5 pc with a turbulent
forcing that mimics star  formation.  They found an important fraction
($\simeq 50 \%$) of thermally unstable gas in their simulation.

Kritsuk \& Norman (2002a, 2002b) have performed 3D simulations
of  HI with a  thermal forcing that mimics  the random fluctuations of
the heating rate derived  by Parravano et al.  (2002). Their aim is to
show  that  interstellar  turbulence   may be   generated by   thermal
instability.

Finally, Piontek \& Ostriker (2004) have recently performed 2D
simulations with the aim of studying the development of the
magneto-rotational instability as well as the thermal instability
in the magnetised warm atomic interstellar gas.

\subsection{Outline of the paper}
In this  paper we further  investigate the dynamical properties  of HI
gas by the mean of 2D numerical simulations.  We explore the behaviour
of  HI  focusing  on  the  formation of  cold  structures  during  the
collision  between  two streams  (converging  flow)  of  WNM.  We  put
special attention to the structure morphology, their internal velocity
dispersion and to  the fraction of gas in the cold  and warm phase and
in an intermediate thermally unstable state.  Our aim is to understand
how those features depend on the turbulence of the flow.

In the second section of  the paper we present the equations, describe
the  thermal  processes, the  numerical  scheme  and  the initial  and
boundary conditions.  We also discuss the drastic numerical resolution
which is  needed in order  to properly describe  the HI flow.   In the
third section we present our  numerical results for a large scale flow
which is weakly or very  turbulent and present statistical analysis of
the simulations.   In the fourth section we  emphasize the correlation
between  turbulence and the  fraction of  thermally unstable  gas.  We
then  develop  a  semi-analytical  model to  understand  the  physical
mechanisms  which  are  responsible  for this  phenomena.   The  fifth
section summarises our results and concludes the paper.

\section{Equations, numerical requirement and initial conditions}
\label{condini}
\subsection{Equations and notations}
In this paper  we consider the Euler  equations for an optically  thin
gas.  The   gas is able  to  cool radiatively and  is  heated up by an
external radiation  field.  The equations  governing the  evolution of
the  fluid are the    classical equations of  hydrodynamics, where   a
cooling function is added in the energy conservation equation:

\begin{alignat}{4}
\label{mcons}
\partial_t \rho       & \ +   \nabla . [\rho u]              & =  & 0 \\
\label{momcons}
\partial_t \rho u     & \ +   \nabla . [\rho u\otimes u + P] & =  & 0 \\
\label{econs}
\partial_t  E         & \ +   \nabla . [u(E + P)]            & =  & - {\cal L}(\rho,T)
\end{alignat}
where $\rho$ is the mass density, $u$ the  velocity, $P$ the pressure,
$E$ the  total energy and  ${\cal L}$  the  cooling function (see next
section).  The gas is  assumed to be a perfect  gas with $\gamma = 5/3$
and with a mean molecular weight $\mu = 1.4 m _H$, where $m _H$ is
the mass of the proton.

The above equations are solved using a second-order Godunov method for
the conservative part. The cooling is applied after the hydrodynamical
step using an implicit scheme and  subcycling when the cooling time is
much smaller than the time step.

\subsection{Thermal processes}
In order to  simulate HI it is important to adopt  a model for thermal
processes which on one hand is  realistic enough and on the other hand
not too computationally expensive.   We have followed closely the work
of Wolfire et al.  (1995,  2003), trying to include only the processes
which are the most relevant to  our study.  The UV flux is taken equal
to  $G _0/1.7$, where  $G_0$ is  the so  called Draine's  flux (Draine
1978).

We   have only kept the   following  cooling  processes, which  are
dominant in the physical conditions of our simulations :
\begin{description}
\item[-] cooling by fine-structure lines of CII (92 K)
\item[-] cooling by fine-structure lines of OI (228 and 326 K)
\item[-] cooling by H (Ly$\alpha$ line)
\item[-] cooling by electron recombination   onto positively charged grains
\end{description}

The heating process is  the photo-electric effect  on small grains  and
polyaromatic hydrocarbons   due   to  the   far-ultraviolet   galactic
radiation.  We have  not taken the heating due  to cosmic rays  and to
the soft  X-rays into account  since these contributions appear  to be
small. Finally   in  order to   calculate  the ionisation  we  use the
approximation proposed by Wolfire et al. (2003).

We have compared  each  term with the   results given in  Wolfire  et
al. (95)   as well as  the  thermal  equilibrium  curve.  Very similar
results have been obtained.

\subsection{Resolution requirement}
The  thermal  condensation of  HI  gas  involves  4 different  spatial
scales, each of them related to a different physical mechanism that we
describe below.

\subsubsection{First scale:  cooling length in the WNM}
The cooling length of the WNM, $\lambda _{\rm cool}$ is defined as the
product of the  cooling time, $\tau _{\rm  cool} = C  _v T / ({\cal L}
\rho)$, by the sound speed, $C  _s$. It represents  the scale at which
the WNM is non-linearly unstable  (Hennebelle \& P\'erault 1999) , i.e 
velocity   fluctuations of spatial   extension  $\simeq  \lambda _{\rm
cool}$ and amplitude $\simeq C  _s$, can undergo a thermal contraction
from WNM  phase into CNM phase.   For typical WNM parameters, $n _{\rm
wnm} \simeq  0.5$  cm$^{-3}$, $T_{\rm wnm}  \simeq   8000$ K, $\lambda
_{\rm cool}$ and is about 10-20 pc whereas for a  higher density of $n  
_{\rm wnm}  \simeq  3 $ cm$^{-3}$, reached  for  example in   shock or
dynamically compressed  layer $\lambda  _{\rm   cool}$ is smaller  and
about 1-3 pc.

\subsubsection{Second scale: fragment size}
The second important  scale of the problem is the  typical size of the
fragments of  CNM.  It is given  by the cooling length  divided by the
density ratio between the CNM and the WNM, $\zeta \simeq 100$. This is
due to the fact that a  piece of initially warm gas contracts until it
reaches  thermal  equilibrium.   Therefore  $\lambda _{\rm  struct}  =
\lambda _{\rm cool} / \zeta \simeq 0.1$ pc. We would like to point out
the fact that  resolving this scale is essential  in order to properly
describe  the structure  of the  cold gas.   Note that  $\lambda _{\rm
struct}$ is the typical size  of the structures in the direction along
which the gas has condensed.  In the other directions, if the collapse
is anisotropic, the scale is  a priori different and is rather related
to the initial  scale of the WNM fluctuation  which has condensed into
the CNM structure.

\subsubsection{Third scale: the Field's length}
The third scale is the Field length (Field 1965) or conduction length,
which is also the typical size  of the front between the 2 phases.  It
is given  by $\lambda _{\rm Field}  = \sqrt{ (\kappa(T) T)  / {\cal L}
}$.  In the WNM, the Field's length is about $10^{-1}$ pc and is about
$10^{-3}$ pc in the CNM.  In  order to have an accurate description of
the thermal front it is necessary to include non-ideal processes, such
as  viscosity  and thermal  diffusion.   Since  the front  propagation
velocity  is proportional  to $\lambda  _{\rm Field}$  (Zel'dovitch \&
Pikel'ner  1969,  Penston \&  Brown  1970),  the  consequences of  not
resolving  properly this  scale  would be  first  to overestimate  the
propagation  of  the thermal  front  and  second  to overestimate  the
fraction  of thermally  unstable gas  located in  the front.   We thus
conclude that in order to simulate properly the HI gas it is essential
that the effective Field length (close to the numerical resolution) in
the  simulation is  small compared  to $\lambda  _{\rm  struct}$. This
ensures that  the growth of  structures due to heat  diffusion remains
small compared to the growth  of structures due to dynamical processes
and that  the fraction  of thermally unstable  gas stabilised  by heat
diffusion  and located  in thermal  fronts  is small  compared to  the
fraction of  cold gas.  The  importance of resolving the  Field length
has been recently carefully analysed by Koyama \& Inutsuka (2004).

\subsubsection{Fourth scale: shocked CNM}
The  fourth  scale  is  a  consequence of  dynamical  processes.   The
internal sound  speed of the  CNM is about  10 times smaller  than the
sound  speed  of  the  WNM.   Consequently, it  is  expected  that  HI
structures experiment  supersonic motions with a  typical Mach number,
$M$, equal to about 10. This may occur through the collision between 2
fragments of  CNM or when  a fragment  of CNM is  swept up by  a shock
wave.   Assuming that  the  polytropic index  of  the CNM  is about  1
(i.e. CNM  is isothermal), this leads  to a density  increase of $\rho
_{\rm shock} = M^2 \rho _{\rm  cnm}$.  The typical size of the shocked
layer is thus:  $\lambda _{shock} = \lambda _{struct}  / M^2$ which is
about $ \simeq 10^{-3}$ pc for $M=10$.  If the numerical resolution is
larger  than $\lambda _{shock}$,  the largest  density reached  in the
simulation  is underestimated  and the  size of  the shocked  layer is
artificially broadened.

As   we  have seen,  simulating the  thermal  condensation  in  HI
satisfactorily, would require to treat spatial scales ranging from few
10 pc to about $\simeq 10^{-3}$ pc simultaneously.

\begin{figure}
\includegraphics[width=8cm]{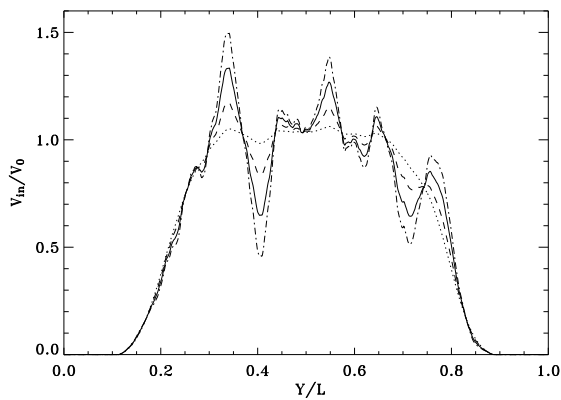}
\caption{The figure illustrates the inflowing velocity field for
$\epsilon=0.5$ (dotted  line), $\epsilon=2$ (dashed line),
   $\epsilon=4$  (solid line) and $\epsilon=6$ (dot-dashed line).  
As can be seen, we have  kept the same phase for the 3
curves.}
\label{Vin}
\end{figure}

\begin{figure*}
\includegraphics[width=15cm,angle=90]{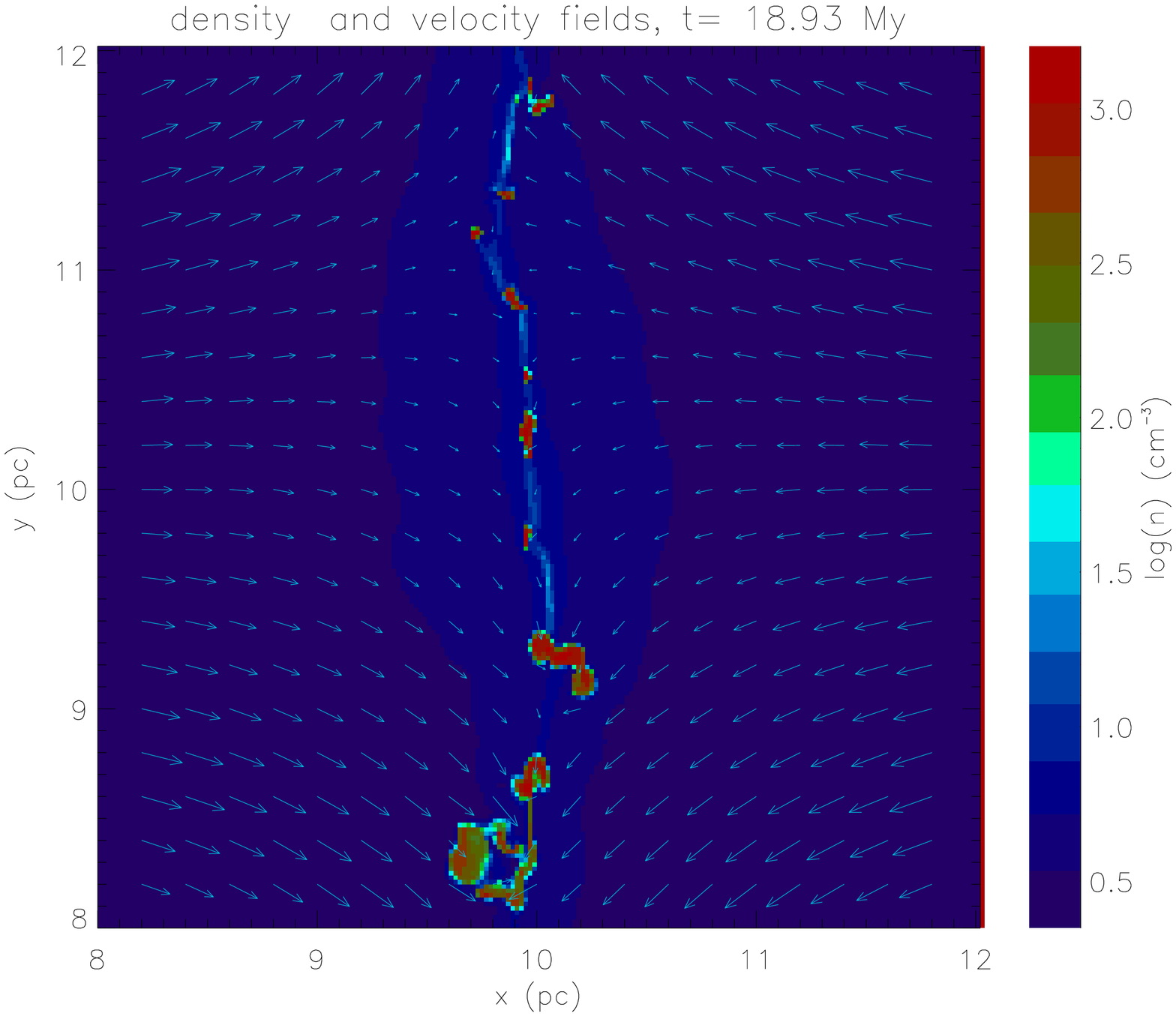}
\caption{Density and velocity fields in the central region 
(4$\times$4 pc$^2$ around the center located at $x=10$ pc $y=10$ pc) of 
the simulation box (20$\times$20 pc$^2$) for $\epsilon=0.5$ 
(weakly turbulent flow) at time t=18.93 Myr. 
The longest arrow represents a velocity of about 15 km/s.}
\label{noturb}
\end{figure*}

\subsection{Boundaries and initial conditions}
Due to the large range  of spatial scales involved in the interstellar
medium it  is not possible  to properly resolve the  scales associated
with the microphysics  (see previous section) and at  the same time to
treat the  large scale flow of  the WNM. It is  therefore necessary to
compromise between the different scales.

Since we are interested in the formation of small dense structures, we
have  chosen  to  resolve  the  small  scales  and  to  use  a  simple
prescription  to model  the large  scale  velocity field  of the  WNM.
Therefore, the simulations used in  this paper are 2D simulations on a
1000$^2$ grid.  In  order to solve the cooling length  the size of the
box is $L = 20 pc$ leading to a numerical resolution of 0.02 pc.  This
numerical  resolution   ensures  that  $\lambda   _{struct}$  is  well
resolved,  and that  it is  larger than  the effective  Field's length
(equal to  about one pixel).   However with this resolution  a shocked
structure is not well described  for Mach numbers larger than about 3.
This  means  that  the  largest densities  reached  during  supersonic
collisions are underestimated in our simulations.

The size of the simulation  box, 20 pc, is marginal  in the sense that
it is not enough to describe the  evolution of a large scale structure
of WNM.  The  large scale flow is  then  imposed by choosing  boundary
conditions that mimic a converging flow  at large scale. The upper and
lower sides of the simulations have free  boundary conditions while on
the left  (resp.  right)  side  the gas is   injected with a  velocity
$V_{in}(y)$ (resp. $-V_{in}(y)$). The density and  the pressure of the
inflowing gas  are chosen such that  the gas is at thermal equilibrium
in the branch  of the WNM phase.  It is thus  thermally stable when it
penetrates in the simulation box.

The function $V_{in}$ describes the velocity of  the inflowing gas and
is given by:
\begin{eqnarray}
\label{inflow_vel}
V_{in}(y) = V_0 \quad e^{-((y-L/2)/0.6L)^6} \quad(1 + U(y)) \\
\nonumber
\mbox{where } \quad U(y) = \epsilon \sum c_i \cos(k_i y + \alpha_i)
\end{eqnarray}
This velocity  field represents a  stream   of gas  with a  transverse
spatial extension of  about $L/2$ and an  average velocity equal to $V
_0$. In order to study the influence of  turbulence, this field can be
modulated by the function $U(y)$.  $\epsilon$ is the amplitude of this
modulation;  $k_i = 2\pi/\lambda_i$     is the wave  number  and   the
wave-length $\lambda_i$ lies between 2 cells  and L/4. We have chosen
a power  law spectrum of index $-1$  for the modulation. Therefore the
$c_i$ are defined by $c_i \propto  k_i^{-1}$ and $\sum  c_i = 1$.  The
$\alpha_i$ are   random phases which  lie  between  0  and $2\pi$.  If
$\epsilon$ is  small then the flow  remains essentially laminar whereas
it becomes turbulent if $\epsilon$  is greater than 1.  Fig.~\ref{Vin}
illustrates the shape  of the inflowing  velocity  field for different
values of $\epsilon$ ($\epsilon$=0.5, 2 and 4).

The gas injected into the simulation box  is at thermal equilibrium in
the    WNM branch and  has a   constant pressure equal   to  $P = 7 \,
10^{-13}$   erg cm$^{-3}$.   The   gas   temperature and  density  are
respectively $\simeq$7100 K and $n=0.76$ cm$^{-3}$.

We have compared 1D results involving a thermal transition between the
2 phases with   the code used   by Hennebelle \& P\'erault (1999)  and
obtained   very similar results   within    few percents of   accuracy
even though the code used in  this work does  not include viscosity and
thermal conduction.

\section{Results}
\label{result}
\begin{figure}
\includegraphics[width=8cm,angle=0]{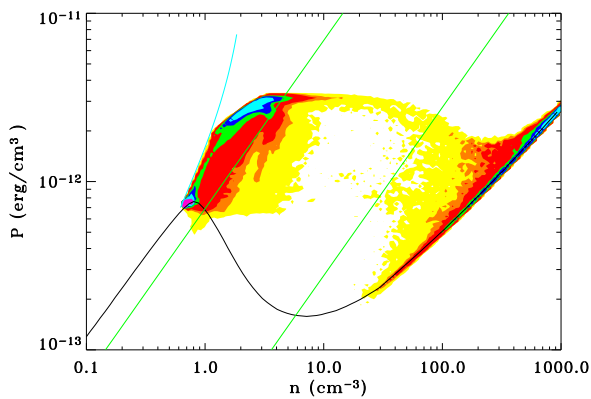}
\caption{Gas  mass fraction  in the  pressure-density diagram  for the
$\epsilon =  0.5$ simulation.  The colours correspond  respectively to
gas mass fraction (in arbitrary units) of (1,5,10,50,100,200,1000) for
(yellow, orange, red,  green, dark blue, light blue,  pink).  The full
black line is the thermal  equilibrium curve.  The green lines are the
isothermal  curve T=5000  K  and T=200  K  and the  blue  line is  the
Hugoniot   curve  of   shocked  gas   corresponding  to   our  initial
conditions.}
\label{press_dens_noturb}
\end{figure}

In  this  section  we  qualitatively  describe results  for  a  weakly
turbulent ($\epsilon=0.5$) and for a very turbulent forcing ($\epsilon
=  4$).  We  then  present a  statistical  analysis for  the 4  cases,
$\epsilon=0.5,  2,  4$ and  6  considering  the  gas fraction  in  the
different phases  and the properties  of the CNM structures  formed in
the simulations.

Initially, there is only warm gas in the simulation box.  Then the two
facing  incoming  flows  generate   a  region of  higher  density  and
pressure in  the center.  This  region is thermally unstable  and cold
structures start  to form.  After some  time (from 5  to 15  Myr), the
simulation reaches a {\it permanent} regime. That is the mass fraction
in  the  different phases,  the  statistical  properties  of the  cold
structures, their  abundance etc... remain constant.   All the results
presented below are given for this {\it permanent} regime.

\begin{figure*}
\includegraphics[width=15cm,angle=90]{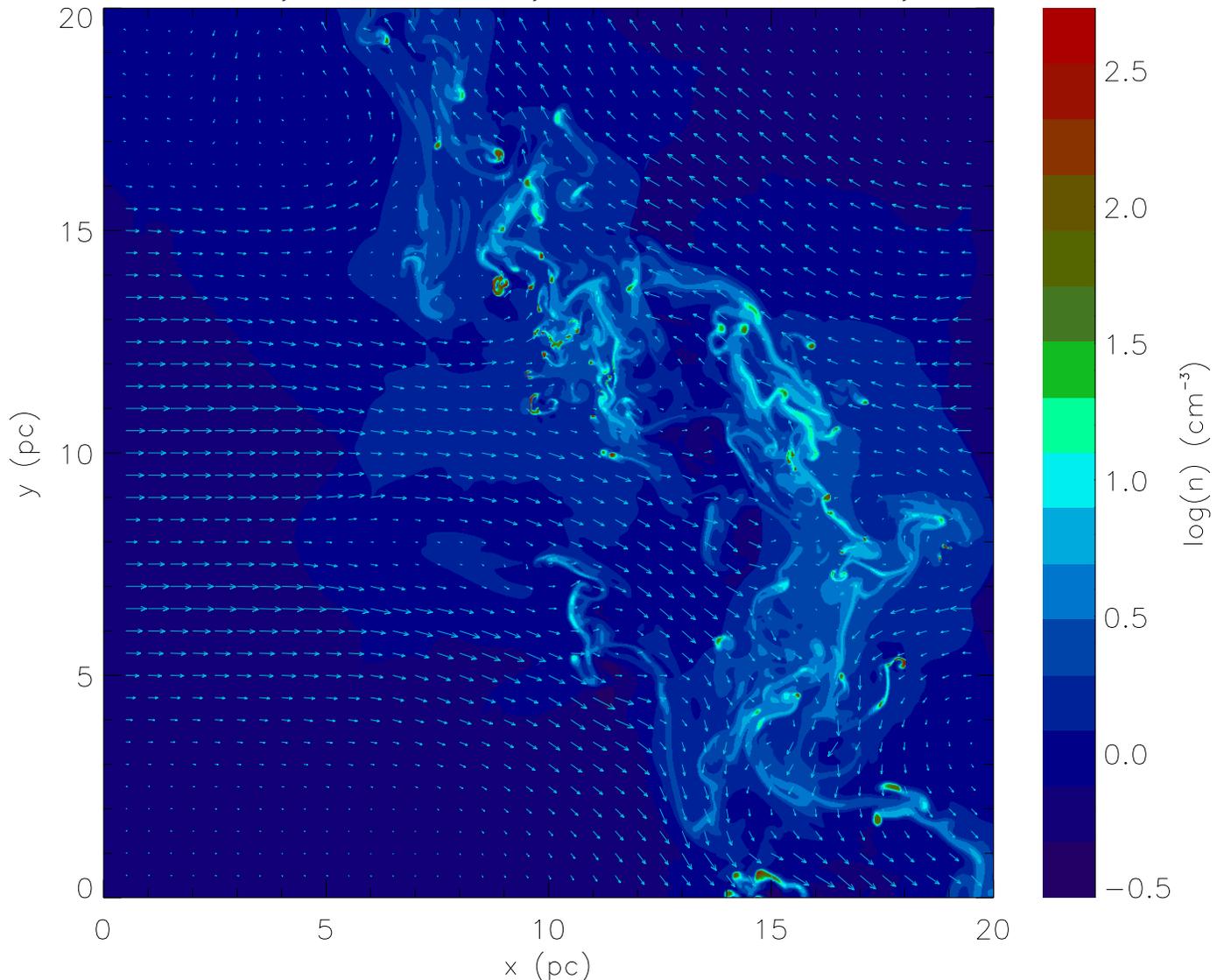}
\caption{Density and velocity fields for a very turbulent forcing
($\epsilon=4$). The longest arrow represents a velocity of about 17 km/s.}
\label{turb}
\end{figure*}

\subsection{Weakly turbulent flow}
When the flow is weakly turbulent,  the ram pressure of the 2 incoming
flows induces the formation of  a high pressure central layer which is
bounded by 2 accretion shocks. This high pressure gas has a density of
about 3  cm$^{-3}$ and a  temperature of about  10$^4$ K.  Due  to the
high  pressure in the  central region  a roughly  homologous expanding
velocity field  develops in the  transverse direction and part  of the
gas escapes the  box continuously.  The velocity field  as well as the
density field in  the central region can be  seen in Fig.~\ref{noturb}
which  displays a  zoom of  4$\times$4  pc$^2$ around  the box  centre
(located  at $x=10$  pc,  $y=10$ pc).   The  layer is  out of  thermal
equilibrium and  its central part becomes thermally  unstable. Part of
the warm  gas condenses into cold  gas. This behaviour  is indeed very
similar to the 1D situation studied by Hennebelle \& P\'erault (1999).
However in the  present case because of the  weak turbulence, the cold
gas layer is unstable and fragments in several parts as can be seen in
Fig.~\ref{noturb} where  about 10  fragments have formed.   Their size
ranges  from about 0.05  to 0.3  pc.  Very  sharp thermal  fronts (2-3
pixels) bound the structures and  connect them to the warm surrounding
medium.  Due  to the high pressure  induced by the  incoming flow, the
density of the  structures is rather high and ranges  from a few $100$
to $\simeq 2000 $ cm$^{-3}$.  The fragments are sometimes connected by
a thin layer of low density gas.

This situation is   very similar to   the 2D numerical simulations  of
Koyama \& Inutsuka (2002) who  study the thermal transition induced by
a  shock  propagating in the warm  phase.  As in  our case,  Koyama \&
Inutsuka (2002)   find that the shocked layer    fragments in few cold
clouds of about 0.1 pc.

Fig.~\ref{press_dens_noturb} gives a  more accurate description of the
thermal state of the gas in  the simulation. It shows the gas fraction
in  the  pressure-density  diagram.   The  full line  is  the  thermal
equilibrium curve. The green lines  are the isothermal curves T=5000 K
and T=200  K and the  blue line is  the Hugoniot curve of  shocked gas
corresponding  to our  initial conditions.   Most of  the warm  gas is
located between the  Hugoniot curve and the isothermal  curve T=5000 K
whereas  most of cold  gas is  very close  to the  thermal equilibrium
curve. There is almost no thermally unstable gas.

\begin{figure*}
\includegraphics[width=15cm,angle=90]{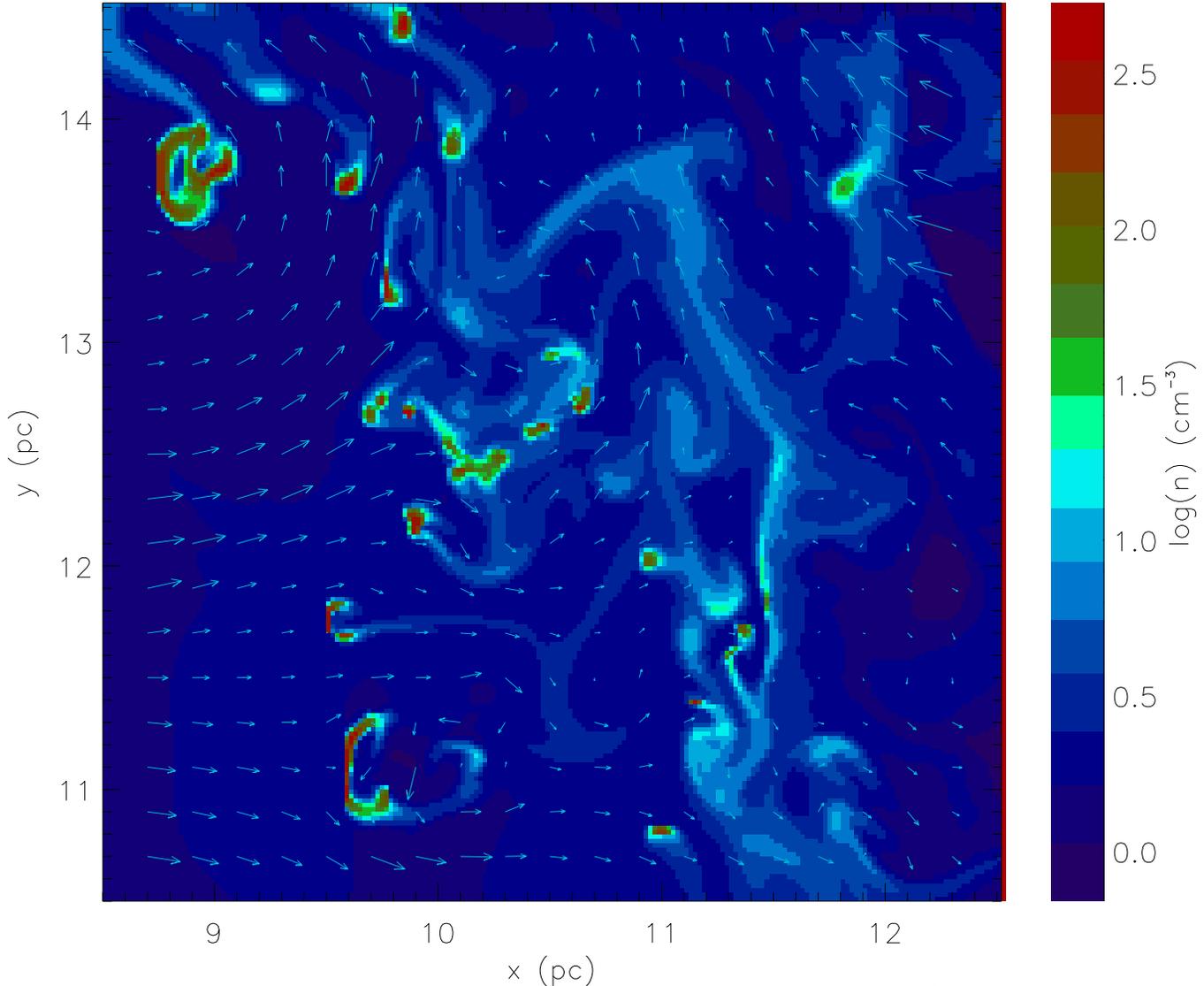}
\caption{Spatial zoom of Fig.~\ref{turb}.
The longest arrow represents a velocity of about 13 km/s.}
\label{turb_zoom}
\end{figure*}

\begin{figure*}
\includegraphics[width=15cm,angle=90]{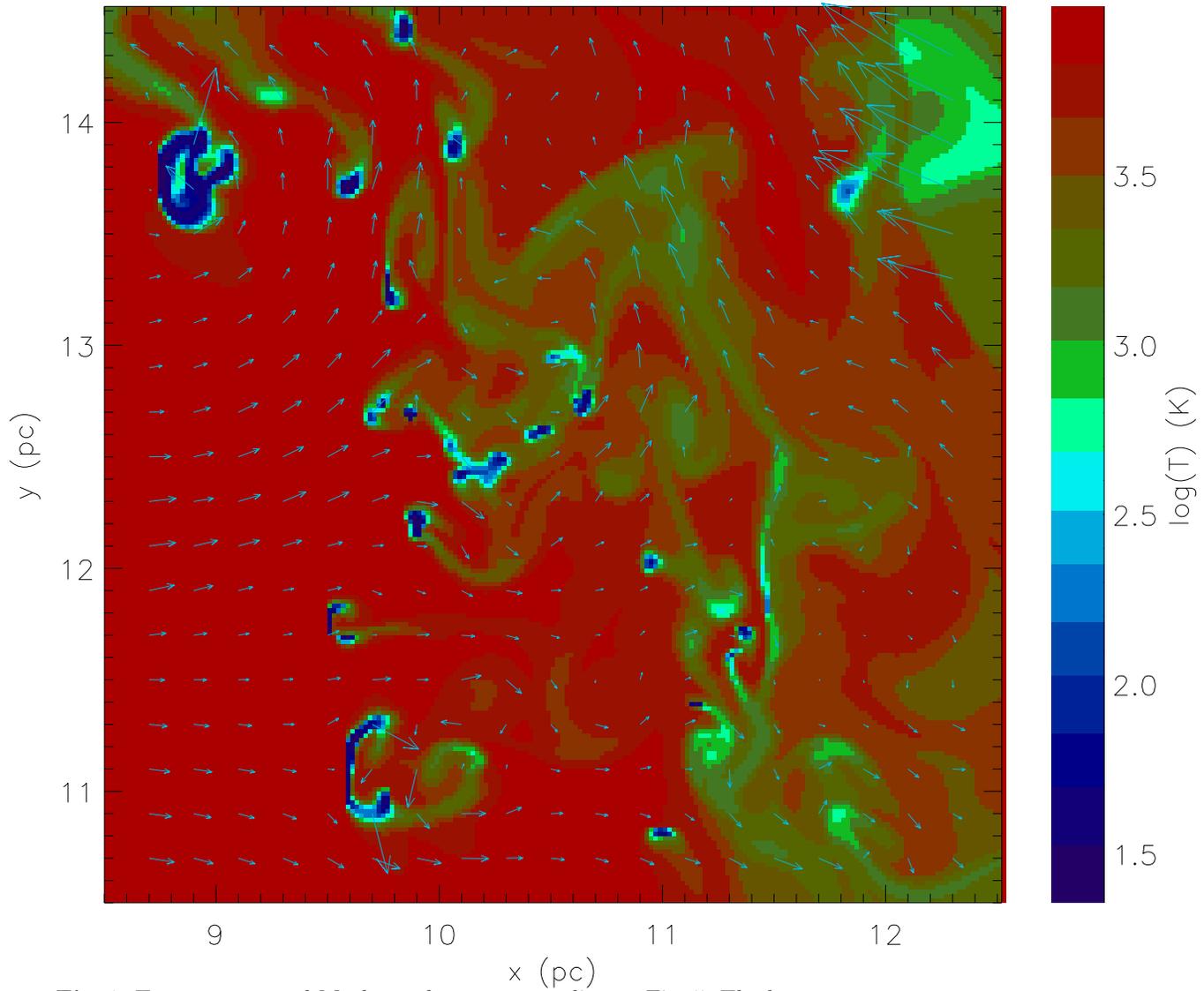}
\caption{Temperature and Mach number corresponding to
Fig.~\ref{turb_zoom}.
The longest arrow represents a mach number of about 10. }
\label{turb_zoom_temp}
\end{figure*}

\subsection{Very turbulent flow}
\begin{figure}
\includegraphics[width=8cm,angle=0]{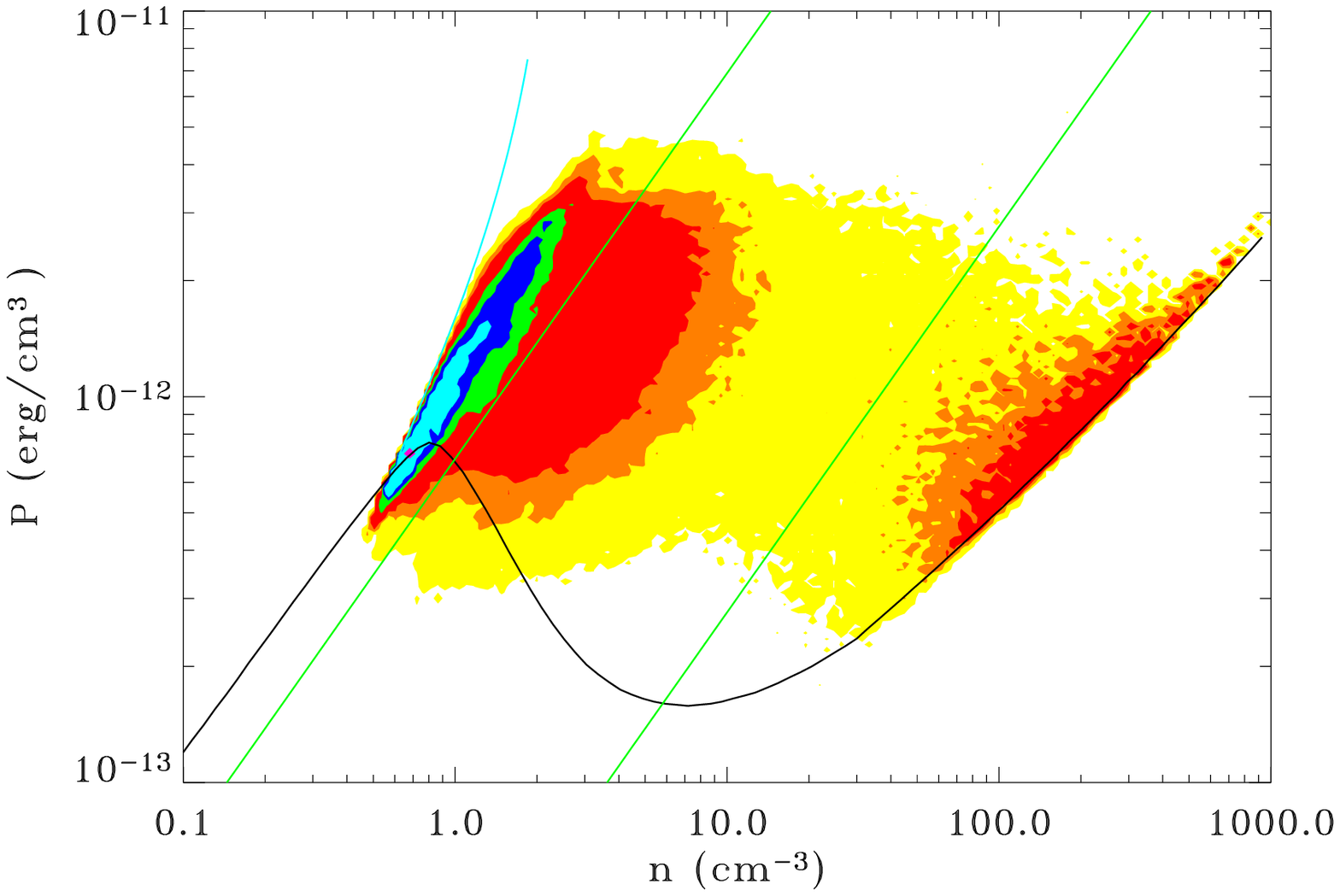}
\caption{Same as Fig.~\ref{press_dens_noturb} for $\epsilon=4$.}
\label{press_dens_turb}
\end{figure}

Fig.~\ref{turb} displays the density and  velocity fields in the whole
box for the very turbulent forcing ($\epsilon=4$) at time t=18.93 Myr.
 The stiff shear
of the gas that penetrates in the box generates turbulence quickly.

The interface between the 2  flows is very irregular and significantly
distorted. The density field is much more complex than in the previous
case. As  illustrated in Fig.~\ref{press_dens_turb},  less material is
at a density of more than few 100 cm$^{-3}$ and a significant fraction
of  the  gas  is  at  intermediate  density  around  5  cm$^{-3}$  and
temperature smaller  than 5000 K,  i.e in a thermally  unstable state.
Moreover,  as can be  seen by  comparing Figs.~\ref{press_dens_noturb}
and \ref{press_dens_turb}  the average  thermal pressure is  higher in
the weakly  turbulent case. This is  due to the fact  that the kinetic
energy  of  the  incoming  flow  (which  is  almost  constant  is  all
simulations)  is converted  into both  thermal pressure  and turbulent
motions.  Therefore,  if the turbulent  motions are high,  the thermal
pressure should be lower.

As can be  seen in Fig.~\ref{turb_zoom} and~\ref{turb_zoom_temp} which
displays  respectively  the  density   and  velocity  fields  and  the
temperature field and Mach number of Fig.~\ref{turb} between $x=8$ and
12 pc  and $y=10.5$ and  14.5 pc, the  thermally unstable gas  is very
filamentary and presents a complex structure (We use the word filament
for the elongated  structures seen in our 2D  simulations. In 3D these
could  become  either sheets  or  filaments,  though  we believe  that
filaments are more likely since  sheets would be more easily broken by
turbulent motions).\\

Several interesting  features appear. The different  phases are highly
interwoven with pockets of warm  gas embedded into filaments of cooler
gas.  This is particularly obvious in Fig.~\ref{turb_zoom_temp} around
$x$ $\simeq$11.5 pc and $y$$\simeq$11.5 pc.

In  spite of  the  presence of  this  unstable gas,  the sharp  fronts
separating the cold  and warm phases are still obvious  as can be seen
for  the  structure  located  at  $x=$9  pc  and  $y=$13.8  pc.   This
relatively  unsurprising result means  that even  in a  very dynamical
medium  the 2-phase  behaviour is  not  suppressed and  may indeed  be
locally rather similar to the standard equilibrium 2-phase model.

Sharp transitions can be seen between the warm phase and the filaments
of thermally  unstable gas as  well (see for  example the front  at $x
\simeq 11.2 \, , \, y \simeq 11.3$ pc).

There  are  more    cold  structures   than  in  the     previous case
($\epsilon=0.5$) but they are  relatively  less dense.  This  suggests
that turbulence   promotes  the fragmentation  of   thermally unstable
medium and  that due to  the more random   motions (i.e less organised
than for the case $\epsilon=0.5$) the average thermal pressure (due to
the high ram pressure) of the medium is reduced.

The cold structures are clearly associated with the unstable gas since
they are  often linked   to other  cold structures   by a filament  of
unstable gas and sometimes embedded in such a filament.

Note that  the size of  the smallest cold  structures is close  to the
mesh   of   our  simulation.    This   means,   as   pointed  out   in
sect.~\ref{condini}, that numerical convergence is clearly not reached
for   those  small   structures.    Moreover  as   can   be  seen   in
Fig.~\ref{turb_zoom},  the  cold  fragments  have a  complex  internal
structure which is also smoothed out by the lack of resolution. \\

We would like  to stress the fact   that in the dynamical regime,  the
2-phase behaviour  is clearly not erased  and that the  description of
the medium as  a continuum of phases   would not be correct.  There is
thermally  unstable gas but it  is highly structured, very filamentary
and connected   to the  denser thermally   stable gas  which   is very
differently  structured.  This means that  new   phenomena due  to the
thermal nature of the flow rather than simple disappearance of the two
phase behaviour occur in this regime.

\subsection{Gas fraction in the different phases}
\begin{figure*}
\includegraphics[width=13cm]{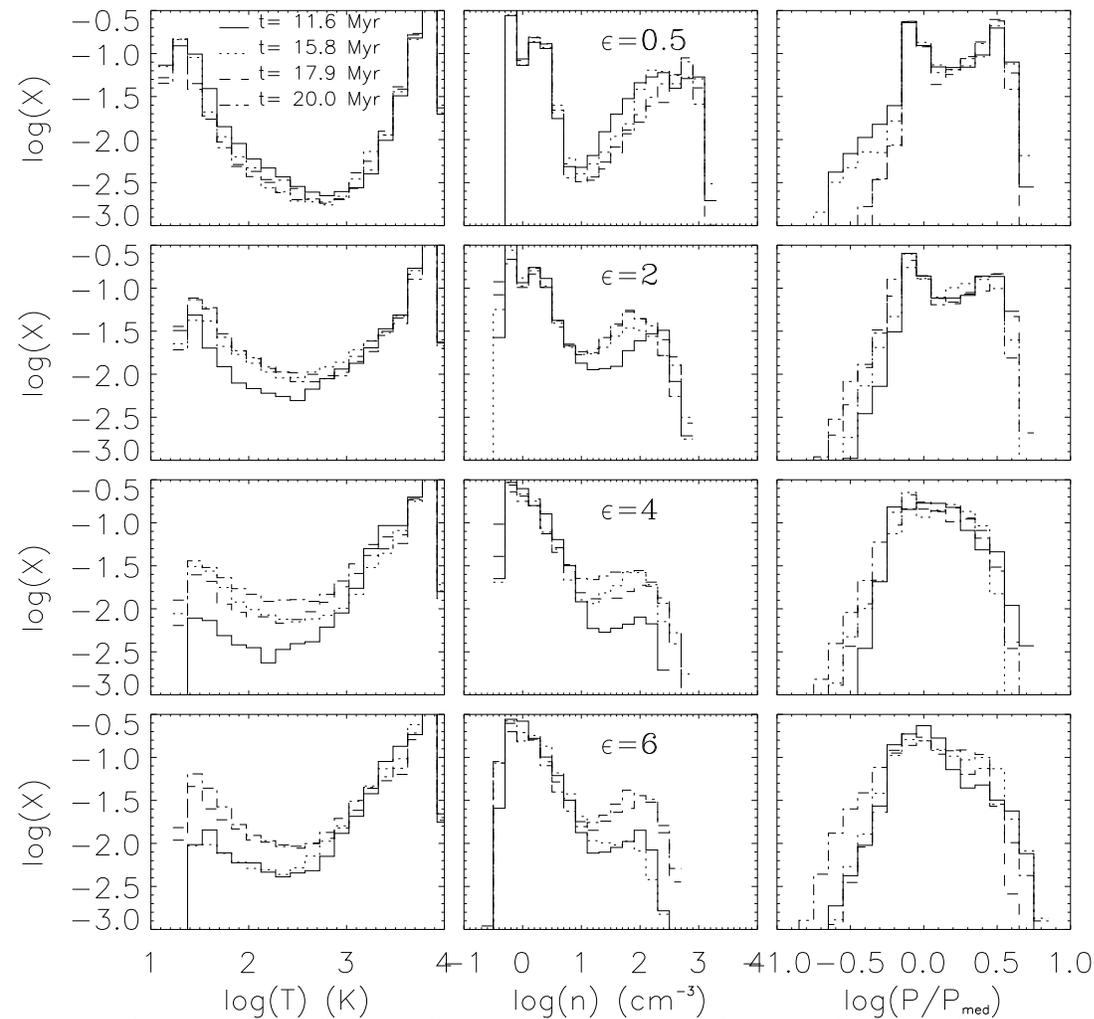}
\caption{Fraction of gas as a function of temperature, density and
pressure for $\epsilon=0.5, \, \epsilon=2$, $\epsilon=4$ and
$\epsilon=6$  at 
times t=11.6, 15.6 and 20 Myr.}
\label{hist_phase}
\end{figure*}

In  this section,  we analyse  the temperature,  density  and pressure
distributions   obtained   in  the   2   cases  presented   previously
($\epsilon=0.5, 4$) and 2  additional cases ($\epsilon=2$ and 6).  For
simplicity  and in  order to  give  simple trends,  we call  thermally
unstable  gas  the gas  having  temperature  between  5000 and  200~K,
whereas the gas with a temperature below 200~K is defined as cold gas.

Fig.~\ref{hist_phase} shows histograms of the fraction of gas ($X$) as
a function  of temperature,  density  and pressure in the four cases
$\epsilon$=0.5, 2, 4 and 6  for 4 time steps:
$t=11.6, \, 15.8$, 17.9 and 20 Myr. 

For the  case $\epsilon=0.5$,  the distributions at  the 4  time steps
presented  are very  similar,  which  means that  the  \pr is  reached
quickly. The  fraction of  thermally unstable gas  is low  ($\simeq 10
\%$) and the  fraction of CNM is about 30$\%$. There  is almost no gas
at  intermediate density  ($n  \simeq 10$  cm$^{-3}$)  whereas $X$  is
almost constant for  density between 100 and 1000  cm$^{-3}$.  It then
drops stiffly for $n >  1000$ cm$^{-3}$.  The pressure fluctuates over
1 order of magnitude. The peak at log$(P / P _{\rm med}) \simeq -0.1 $
is due to  the preshocked gas whereas the peak at  $\simeq 0.5$ is the
postshocked gas ($P_{\rm med}$ is the median pressure).

In the case $\epsilon =2$, the  fraction of cold gas is slightly lower
(20$\%$)  than for  $\epsilon=0.5$ whereas  the fraction  of thermally
unstable gas is larger (20$\%$).  The fraction of gas at density above
10 cm$^{-3}$ is lower at the first time step than later on which means
that the \pr is longer to reach.

For the cases $\epsilon=4$ and  6, these trends  are even clearer.  It
is seen  that the distribution at time  $t=11.6$ Myr is significantly
different from  the distributions at time  $t=15.8$, 17.9 and 20 Myr.
The fraction of dense gas is smaller at  $t=11.6$ Myr by approximately
a factor 3.  This means that the \pr is much  longer to reach than for
the previous cases.  Anticipating the mechanism that will be discussed
in Sect.~\ref{turb_therm}, we interpret this result as the consequence
of the fact   that turbulence  is  able  to stabilise  the   thermally
unstable gas making it  able to last  longer and consequently delaying
the formation of the  cold gas.  In the same  way, it is seen that the
fraction of cold gas when statistical equilibrium is reached ($t$=17.9
and 20 Myr) is about $10 \%$ and thus significantly lower than for the
cases  $\epsilon=0.5$  and  2.   On  the contrary,   the   fraction of
thermally unstable    gas is higher (about   $30\%$)  and the  drop at
intermediate density ($\simeq 10$ cm$^{-3}$) is less clear.

\subsection{Analysis of the CNM structures}
\subsubsection{The analysis}
\label{analysis}
We  now investigate  the  properties  of the  CNM  structures in  more
details. We compute their surface and aspect ratio, their internal and
relative  (to each  other) velocity  dispersion, as  well  as internal
density, temperature and pressure distributions.  A precise definition
of a "structure"  is needed. We define a structure as  a connex set of
pixels having a density above a  given threshold, $n_s \, = \, 30$
cm$^{-3}$.   This definition has  a clear  physical meaning  since the
front that connects the 2 phases is stiff and therefore the structures
do not  depend very much on  the arbitrary threshold  $n_s$.

Once having  identified the structures, we compute  the inertia matrix
$I$ defined by  $I _{xx}=\int \rho x  ^2 dx dy$, $I _{yy}  = \int \rho
y^2 dx dy$ and $I _{xy}= I_{yx} =  \int \rho x y dx dy$. It admits the
2 eigenvalues  $\lambda _1 \ge \lambda  _2$.  The aspect  ratio $r$ is
then defined by $r = \sqrt{\lambda _2/ \lambda _1}$.  We also consider
the velocity of  the structure, ${\bf V} _{\rm  struct}$, its surface,
$S$, average  density, $<\rho>$, temperature,  $<T>$, and internal
dispersion velocity defined as:
\begin{eqnarray}
\delta V ^2 = {1 \over N}
\sum _{i=1,N} {\rho _i \over <\rho>} \left({\bf V} _{\rm struct} -  {\bf v}
_i \right)^2 , 
\end{eqnarray}
where $N$ is the number of pixels in the structure.

Last, we  compute  the  temperature, density and pressure variance
for  a structure defined as:
\begin{eqnarray}
\delta T = \sqrt{ {1 \over N} \sum  _{i=1,N} {\rho_i \over <\rho>} \left(1 -  {T _i \over <T> } \right)^2 }  
\end{eqnarray}

\subsubsection{Morphology and properties of the CNM}

\begin{figure}
\includegraphics[width=8cm]{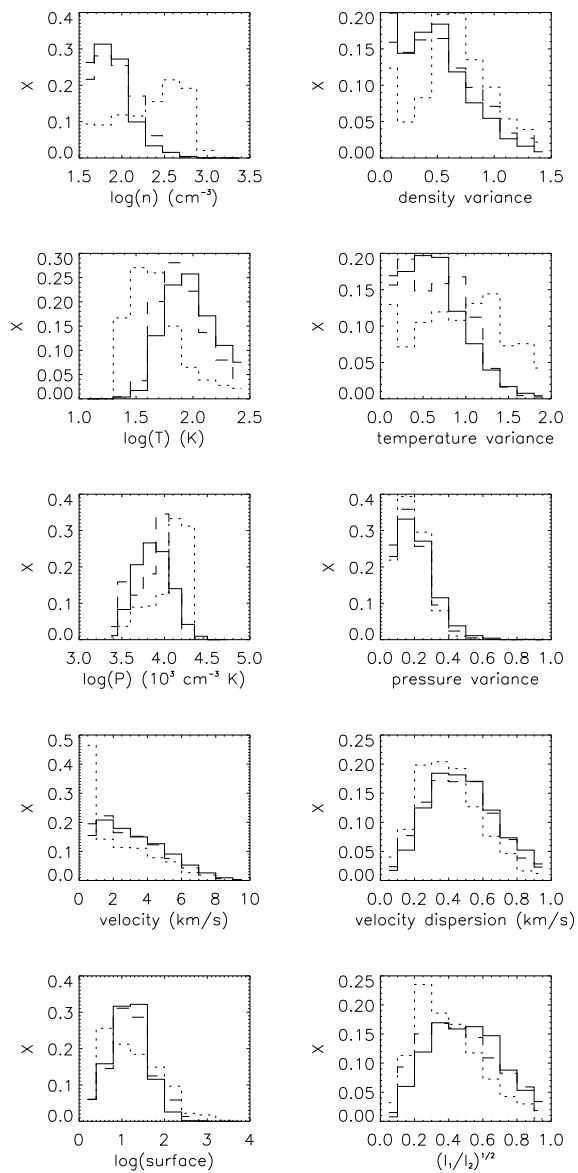}
\caption{Properties of CNM structures for the 3 cases
 $\epsilon=0.5$ (dashed  line),  $\epsilon=2$ (dot-dashed   line)  and
 $\epsilon=4$   (full  line).   The histograms  display  the  density,
 temperature, pressure, velocity, distributions as well as their
variance, the
 size  (in  pixels) and the aspect   ratio  distribution. See text for
 definitions of these quantities.}
\label{hist_cnm}
\end{figure}

Fig.~\ref{hist_cnm}   displays  the   distribution  of   the  density,
temperature, pressure and velocity of  the structures as well as their
variance  (see  Sect.~\ref{analysis} for  the  definitions).  It  also
shows their surfaces (in pixels)  and their aspect ratio.  In order to
obtain the distributions, we have  added the structures obtained in 20
different time steps (separated by about 0.5 Myr in time) leading to a
total number of about 800 to  1500 structures of CNM (depending on the
value of $\epsilon$).

The density histograms confirm  the trend  mentioned in the  previous
section that the structure are much denser in the weakly turbulent
 case than in
the  turbulent one  by    approximately   a  factor  10.   They   are
consequently colder (factor  3) and  have  a higher thermal  pressure
(factor 3) for $\epsilon=0.5$ than for $\epsilon=4$.

The  density and  temperature  variances (respectively  $\simeq$0.6 and
$\simeq$0.7  for   $\epsilon=4$  and   $\simeq$0.7 and   $\simeq$1 for
$\epsilon=0.5$)  indicate  that   these  quantities vary  significantly
inside one structure around their mean value. The pressure variance is
significantly lower ($\simeq 0.3$) which  indicates that the structures
are on average not far from mechanical equilibrium.

The structure velocity  ranges from 0 to 8  km/s and is slightly lower
in the  weakly turbulent case  ($\epsilon=0.5$) than in  the turbulent
one.  This  is  consistent with the  results  obtained  by Koyoma \&
Inutsuka  (2002) who found  a comparable velocity dispersion for their
clouds. The internal velocity dispersion is also slightly lower in the
weakly turbulent  case, for  which it  peaks  at  $\simeq  0.35$ km/s,
whereas   it  peaks at $\simeq    0.45$  km/s for  $\epsilon=4$.  This
indicates that most of the internal motions  are subsonic with respect
to the  internal sound speed  ($\simeq$0.8 km/s) of the CNM structures
and is consistent with the small pressure variance.

Finally, it  is seen that the  mean surface of the  structure is about
20-30 pixels wich gives a typical length of about $\simeq 5$ pixels or
$\simeq$0.1 pc.  As already mentioned the structures  are smaller when
turbulence is higher. The structures have a mean aspect ratio of about
0.5 for $\epsilon=4$ and 0.35 for $\epsilon=0.5$.

\section{Turbulence and thermally unstable gas}
\label{turb_therm}
In this section we  present a more  detailed analysis  of the  role of
turbulence   in  producing the filaments    of  thermally unstable gas
discussed in the previous section.  We first present a semi-analytical
model that describes the transition  of a gas  clump from the warm  to
the  cold phase  and  propose an explanation   for  the origin  of the
unstable gas.   We then compare the  predictions of the model with the
numerical results.

\subsection{A semi-analytical model for the evolution of a fluid particle}

\setlength{\unitlength}{1cm} 
\begin{figure}
\begin{picture}(0.,10.)
\put(0.,3.84){\includegraphics[width=7cm]{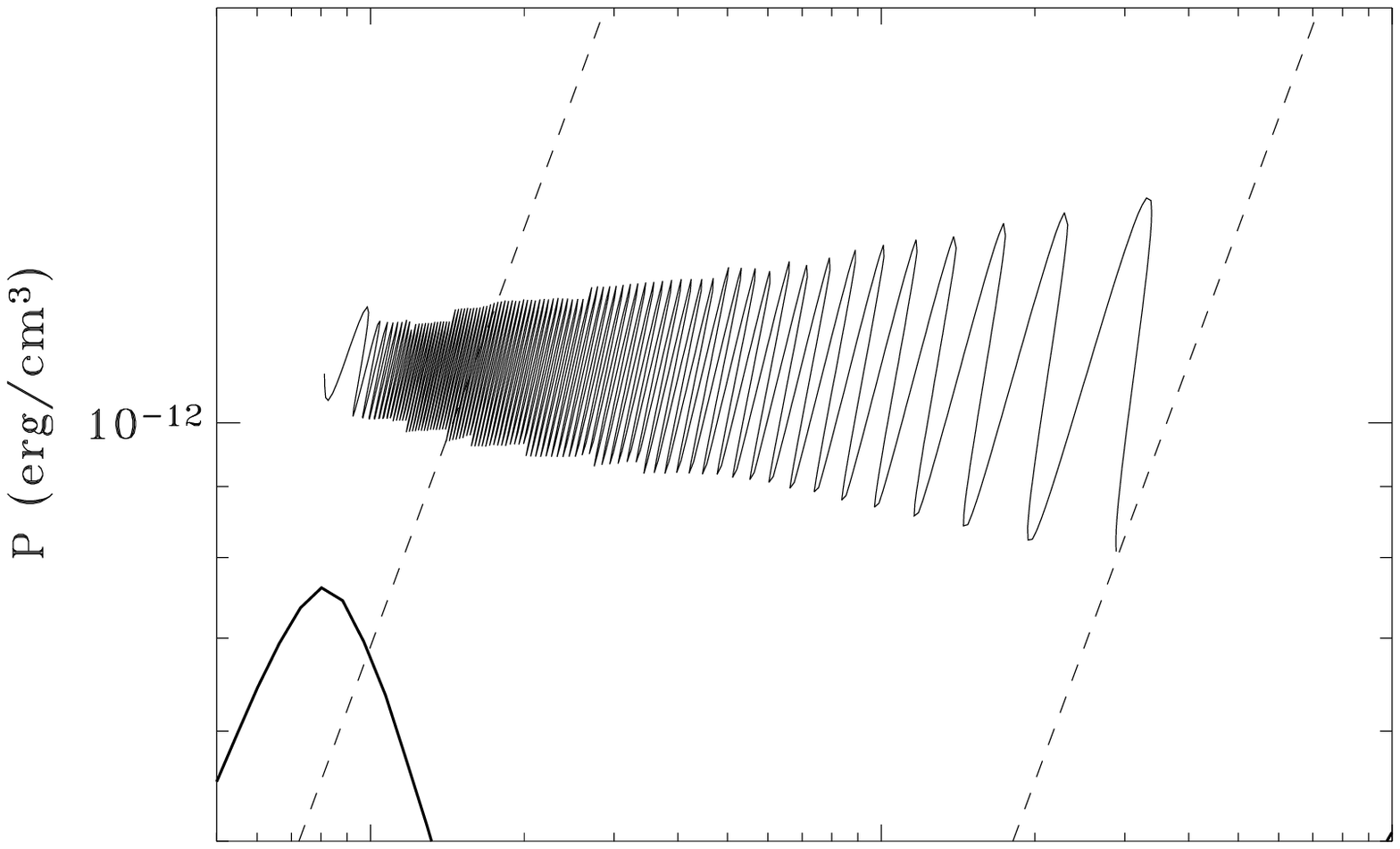}}
\put(0.,0.){\includegraphics[width=7cm]{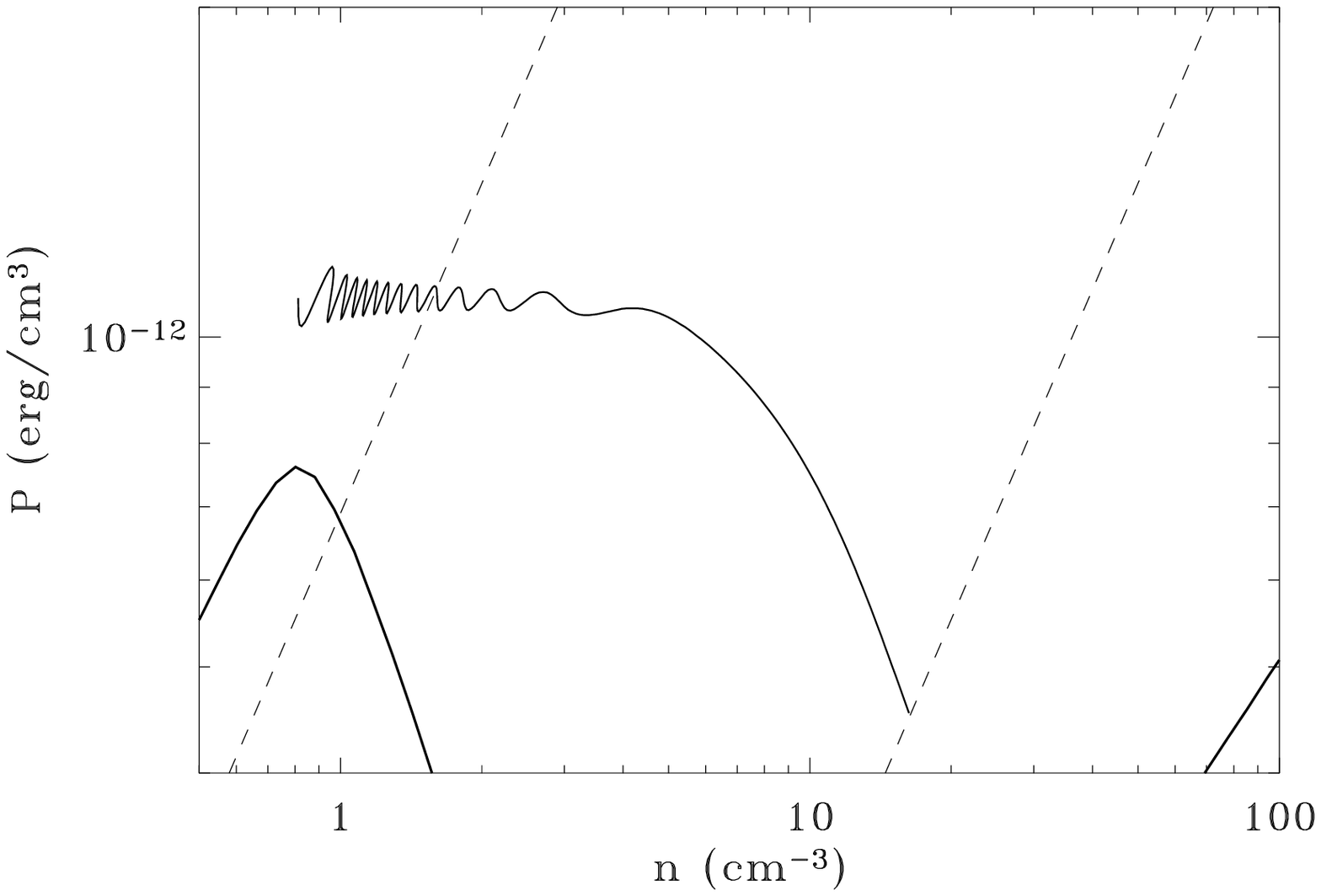}}
\end{picture}
\caption{Evolution of the semi-analytical model for 
a model with $(\theta,\Sigma) = (0,0)$ (bottom panel) and for the same
model  with $(\theta,\Sigma) =   (0,0.8/\tau _{cool})$  (Upper  panel).
The dashed lines are the $5000$~K and $200$~K isothermal curves.}
\label{res_model}
\end{figure}

\subsubsection{Formalism and physical interpretations}
In order to   understand how turbulence generates   thermally unstable
gas, we have developed a semi-analytical model  which is presented in
the appendix.  The underlying physical idea of  the model is simply to
follow a  fluid element and to  calculate the evolution of its density
and temperature by computing  the cooling and heating  of the gas  and
its  geometrical  evolution.   It  is based   on a  Lagrangian form of
Eqs.~(\ref{mcons})-(\ref{econs}) and the  assumptions  that  the fluid
element can  be  simply described  by  its  semi-minor  and major axis
length $a(t)$ and $b(t)$,  its internal pressure $P(t)$, the  external
pressure  $P  _{\rm  ext}$ and   its   density $\rho(t)$.  With  these
assumptions  the   fluid element can   be  described by  the following
equations:
\begin{eqnarray}
\label{equ1}
\dot{\rho} + & \rho \theta & = 0, \\
\dot{\cal{E}} + & P / \rho \theta & =  - \frac{1}{\rho} {\cal
L}(\rho,T), \\
\dot{\theta}   + & \theta^2 / 2 + 2 \Sigma^2 & \simeq - \frac{
P_{\rm ext} - P(t) }{\rho(t)} \left( {1 \over a(t)^2} + {1 \over
b(t)^2} \right), \\
\dot{\Sigma} + & \theta \Sigma & \simeq -\frac{1}{2} 
\frac{P_{\rm ext} - P(t) }{\rho(t)} \left( {1 \over a(t)^2} - {1 \over
b(t)^2} \right).
\label{equ2}
\end{eqnarray}

In these  equations $\theta$ is the divergence  of the velocity field,
$\Sigma= 0.5(\partial_x u_x - \partial_y u_y)$ (where the axis x and y
are  the eigenframe of  the shear tensor,  see  appendix), $a(t) = a_0
\exp  ( \int_0^t (\theta/2  +  \Sigma) dt)$  and  $b(t)  = a_0 \exp  (
\int_0^t (\theta/2 -\Sigma) dt)$. Note that $\Sigma$   describes the 
straining motions (e.g.  Acheson   1990) which are divergence  free and
involve stretching and squashing in mutually perpendicular directions.

%
%
There are three characteristic time scales that can be associated to a
collapsing  fluid  element.   The  cooling  time,  $\tau_{cool}$  (see
appendix),  and  two  dynamical  time  scales  :  $\tau_{dyn}^{\pm}  =
-\theta/2 \pm \Sigma$ which correspond  to the collapse time along the
two principal axis  of the fluid element.  In the  case of an isobaric
evolution  (see appendix  for  details), we  have $\tau_{cool}  \simeq
\tau_{dyn}$  and the  characteristic time  of evolution  of  the fluid
element will be the cooling  time.  If one takes the pressure gradient
into account, the different time  scales are not necessarily equal and
more complex situations can arise.

When there is no turbulence the kinetic energy of the incoming flow is
mainly  converted into  thermal pressure  at  the shock  while in  the
turbulent  case  it  is  converted  both  into  thermal  pressure  and
turbulent motion. Consequently,  the shocked gas in  the low turbulence
simulation  has a  very short  cooling  time (because  of their  high
pressure) and a  very long dynamical time because  most of the kinetic
energy  was turned  into thermal  pressure.  On  the contrary,  in the
highly turbulent simulation, the shocked gas has a much longer cooling
time and a smaller dynamical time.

In particular, if  a dynamical time scale is  shorter than the cooling
time, it  means that  an axis  is collapsing faster  than the  gas can
cool.   Therefore the  pressure and  the pressure  gradient  will rise
along this particular direction which will eventually bounce and start
to expand.

In view of  this, it is possible  to identify two  physical mechanisms
which explain why thermally unstable gas is generated by turbulence.

First,    as   illustrated    by    Fig.~\ref{press_dens_noturb}   and
\ref{press_dens_turb}, the thermal pressure is higher in the case of a
weakly turbulent flow than in the case of a very turbulent flow making
the cooling time much shorter in the first case than in the second one.

The second mechanism which enhances the fraction of thermally unstable
gas  is a  dynamical one  related to  the straining  motions.   Let us
consider  for reference  an  initially spheroidal  piece of  thermally
unstable gas which contracts  isotropically.  In such case $\theta$ is
negative and  $\Sigma$ vanishes.   The density, $\rho$,  increases and
because of  the cooling, the internal energy  decreases.  Therefore if
the dynamical time scale is larger than the cooling time, the pressure
$P$ decreases as well and the  piece of gas keeps contracting. In this
case the dynamical times are equal  on both axis.  Let us now consider
the same  piece of  gas but with  a positive $\Sigma$.   The dynamical
time scale will  be reduced on the y-axis and  enhanced on the x-axis.
The cooling  time will  remain the  same (at least  in a  first phase)
since the evolution of the density and of the pressure depends only on
$\theta$.  However, since the rate  of contraction along the y-axis is
higher, the pressure gradient will also be greater and the contraction
can be slowed or even turned into expansion.

All  these dynamical effect  are amplified  by  the thermally unstable
nature of the gas.   If for  any  dynamical reason the  contraction is
reduced, then  the density will be  lower and since $\partial_\rho P <
0$  the pressure  will  be higher and it   will therefore be even more
difficult to collapse.

This    stabilisation   of    unstable   gas    is    illustrated   in
Fig.~\ref{res_model}  where the evolution  of a  fluid element  in the
pressure-density  diagram  is  displayed.   The bottom  panel  is  for
$(\theta,\Sigma)=(0,0)$,    whereas   the    upper   panel    is   for
$(\theta,\Sigma)=(0,0.8/\tau   _{\rm  cool})$.    Since  we   want  to
illustrate the dynamical effect of  $\Sigma$, we have kept the initial
thermal pressure identical  for the two models; but  as was mentioned
earlier, in  the simulation the  initial thermal pressure is  lower on
average when  $\Sigma$ is large.  It  is seen that in  the first case,
the gas contracts rapidly with  almost no oscillations and the thermal
pressure  decreases rapidly.   In the  second case  the  fluid element
oscillated  rapidly and  the evolution  is more  isobaric.   The fluid
particle remains in the unstable domain about twice the time it spends
in the unstable domain when $\Sigma=0$. \\

Finally, we stress that both mechanisms presented to explain the
presence of unstable gas are related to the presence of turbulence. It
is  expected that filamentary   structures  will be  produced by  such
motions.  Since straining motions  (i.e.  $\Sigma$) are produced in  a
turbulent flow,  this mechanism  explains  how turbulence may  produce
filaments of thermally unstable gas.

\subsubsection{Predictions of the semi-analytical model}
\label{predict_semana}
\begin{figure}
\includegraphics[width=7cm]{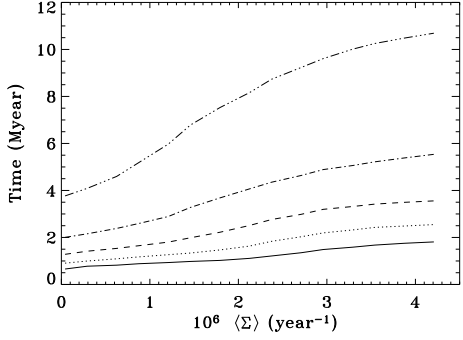}
\caption{ Time spent by the fluid particle
 in the thermally unstable domain (defined as 200  K $<$ T $<$ 5000 K)
 during its  contraction   as a  function of  $\Sigma$.   The 5 curves
 represent respectively  (from top  to bottom) a   mach number of 1.2,
 1.4, 1.6, 1.8 and 2. }
\label{temps_ins}
\end{figure}

We     now     present     the     quantitative     predictions     of
Eqs.~(\ref{equ1})-(\ref{equ2})  for the  time spent  in  the thermally
unstable domain. We consider a fluid  element of warm gas after it has
been  shocked  and  we  follow   its  evolution  until  it  reaches  a
temperature  of  200  K.  We  explore  a  large  ensemble  of  initial
conditions defined  by the initial  state of the gas  (temperature and
density), the initial  values of $\theta$ and $\Sigma$  as well as the
size of the fluid particle, $l$.

Fig.~\ref{temps_ins}  shows  the  mean  time spent  in  the  thermally
unstable domain as  a function of the initial value  of $\Sigma$ for 5
values of the initial mach number,  $M$, namely 1.2, 1.4, 1.6, 1.8 and
2.  The initial  density and temperature are obtained  by applying the
Rankine-Hugoniot relations for an  isothermal shock of mach number $M$
and preshocked gas of density and temperature equal to the values used
as initial conditions in  our simulations i.e $n_0=0.76$ cm$^{-3}$
and $T_0= 7100$  K.  For each value of $M$  and $\Sigma$, we calculate
the  particle  evolution  for   $\theta$  ranging  from  -5.0  to  2.5
Myr$^{-1}$, for  $l$ ranging from  0.5 to 10  pc and compute  the mean
value.

It is seen that, as expected, the time spent  by the fluid particle in
the thermally unstable domain,  $\tau  _{u}$, decreases when  the Mach
number   increases and  increases when $\Sigma$   increases. The ratio
between   these times for   $M=2$    and $\Sigma=0$ and   $M=1.2$  and
$\Sigma=4$ is about 10.

Under simple assumptions, the results presented in Fig.~\ref{temps_ins}
can  be used  to  predict the relative  ratio  between the fraction of
unstable and cold gas  for different values of $M$  and $\Sigma$.  Let
$m  _u$ and $m  _c$    be the mass   of   the unstable and cold    gas
respectively.   The cold gas  forms from the  collapse of the unstable
gas and  disappears when it reaches  the upper or   lower sides of our
computational box. Therefore one has:
\begin{equation}
{ d m _c \over d t } \simeq {m _u \over \tau _u } -  {m _c \over \tau _c } 
\label{equ_moy}
\end{equation}
where $\tau _c$  is the mean  time spent by the cold gas in our
simulation box.
Therefore when statistical equilibrium is reached in our box, we have:
\begin{equation} 
{ \tau _{u} (M _1, \Sigma _1) \over  \tau _{u} (M _2, \Sigma _2) } \simeq 
\left( { m_u \over  m _c } \right) _{(M_1,\Sigma _1)} / 
\left( { m_u \over  m _c } \right) _{(M_2,\Sigma _2)}
\label{equilibre}
\end{equation}
where we have assumed that $\tau _c$ does not vary significantly with 
$M$ and $\Sigma$.

\subsection{Comparison with the numerical results}
\begin{figure}
\includegraphics[width=8cm]{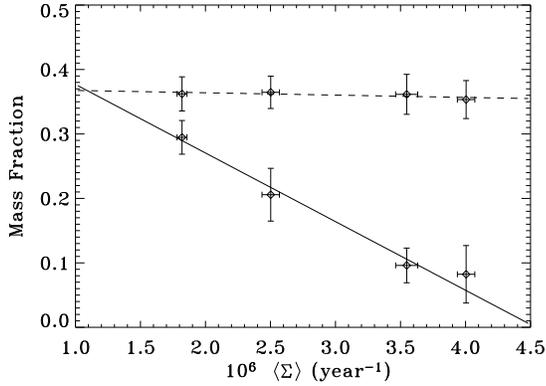}
\caption{Fraction of cold  gas (full line)
and fraction of thermally unstable gas plus cold  gas (dashed line) as
a function of $<\Sigma>$  for the 4  cases $\epsilon=$0.5, 2, 4  and 6
for  various time steps.   Each point  correspond  to the average over
several    outputs of a  simulation with   a given  $\epsilon$ and the
error bars  show  the total dispersion  of  the results  (100\%  of the
points are within the error bar). }
\label{corre_sig_inst}
\end{figure}

\begin{figure}
\begin{picture}(0,12)
\put(0.,8.){\includegraphics[width=7cm]{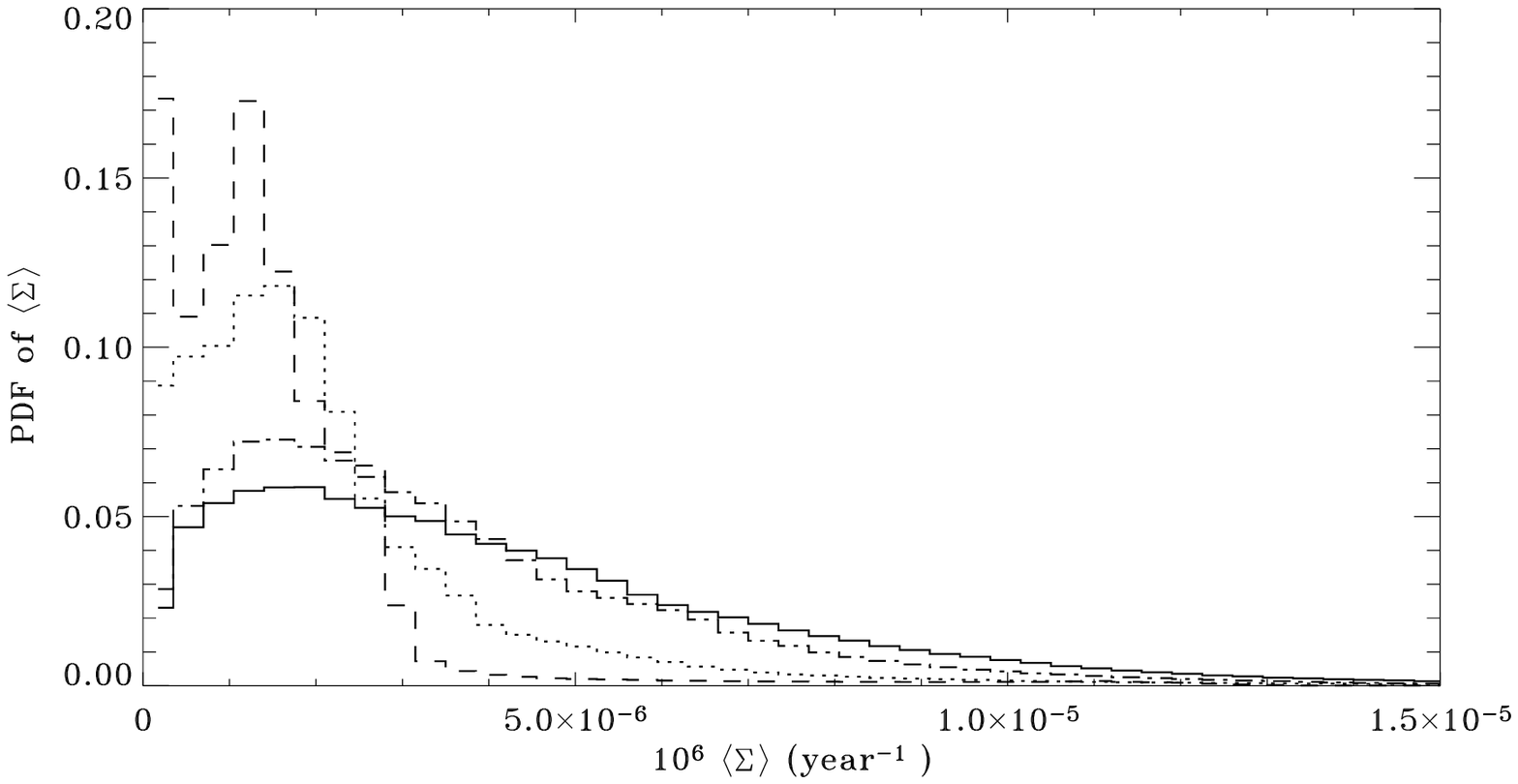}}
\put(0.,4.){\includegraphics[width=7cm]{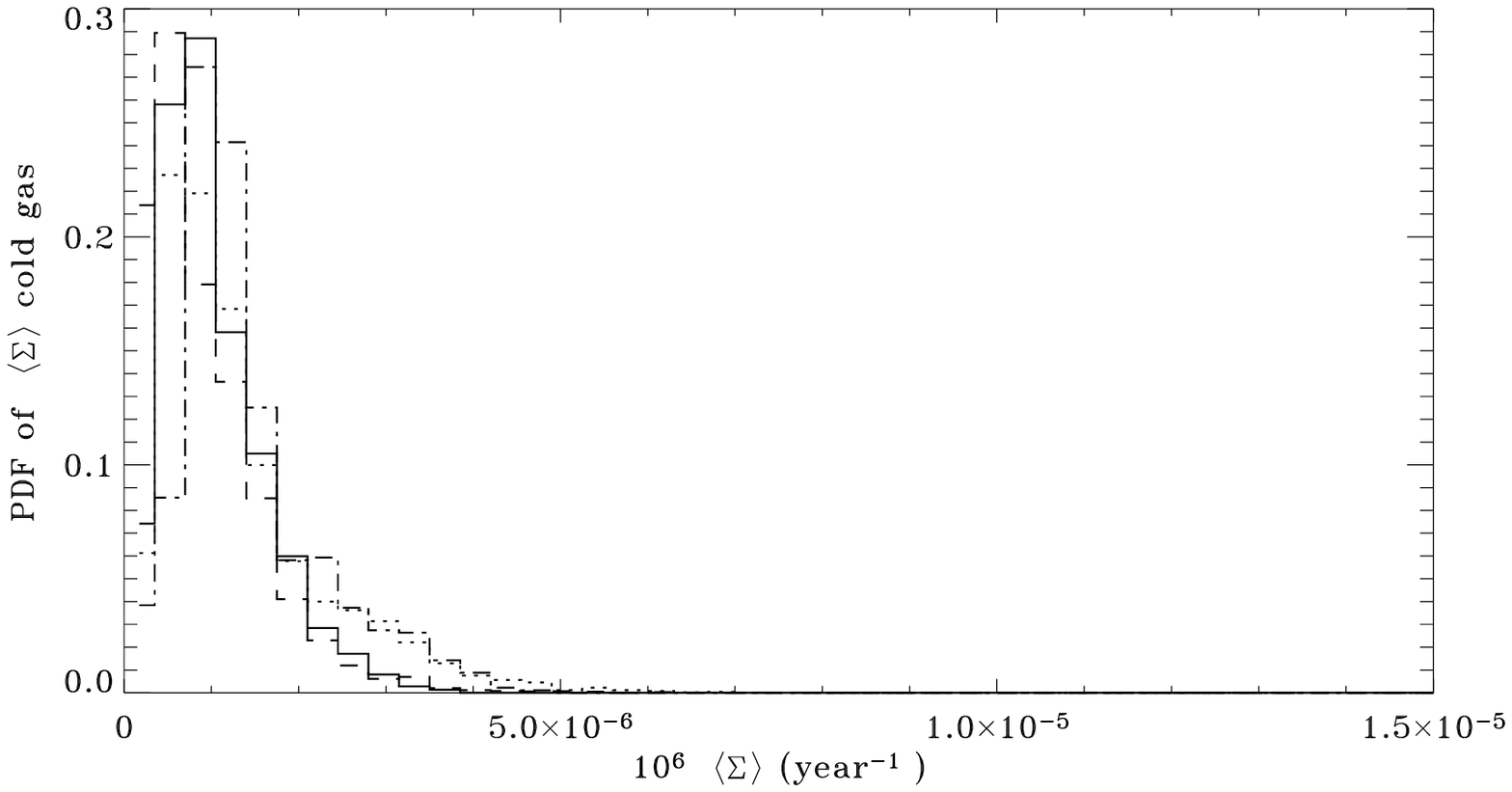}}
\put(0.,0.){\includegraphics[width=7cm]{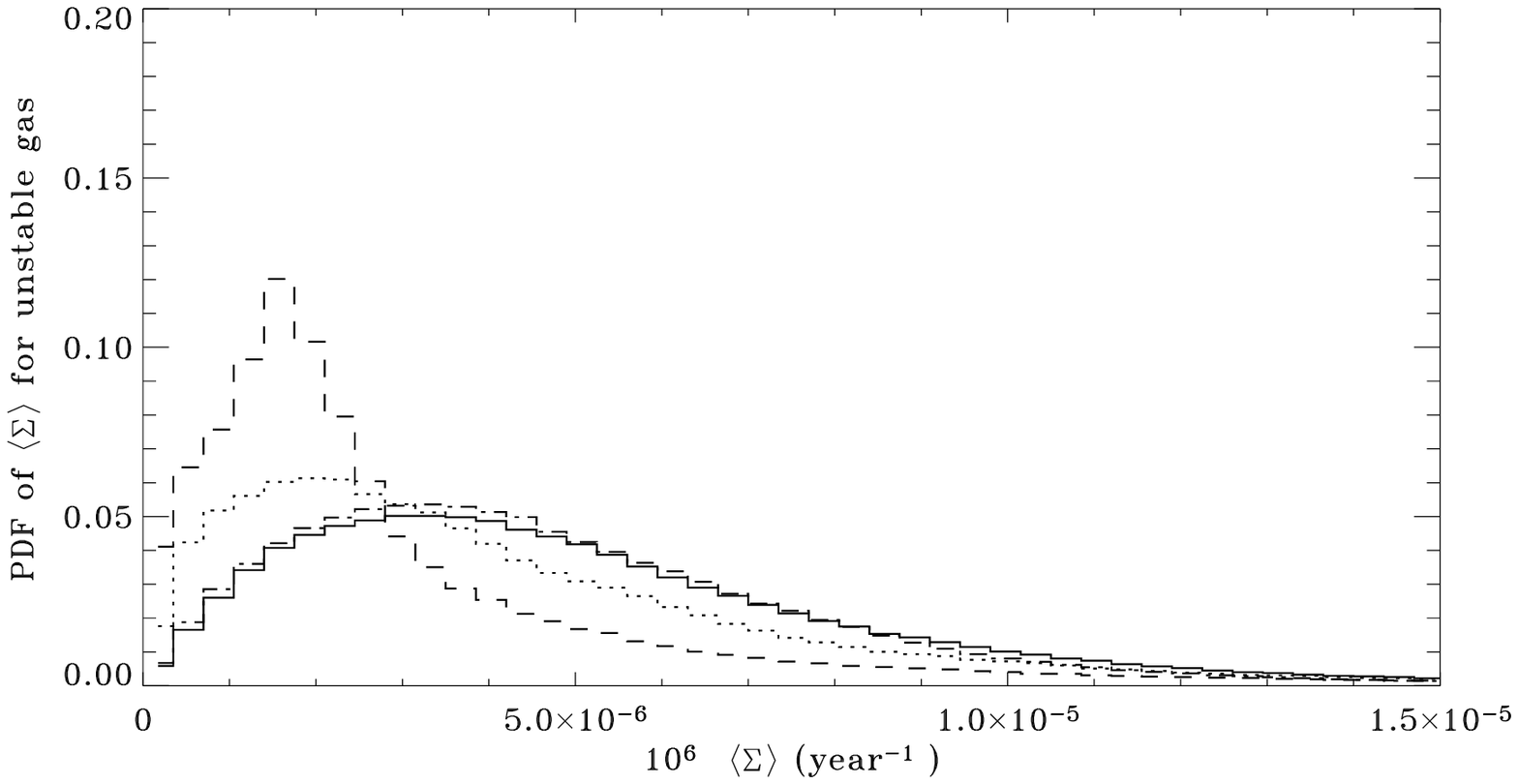}}
\end{picture}
\caption{ Probability distribution function (pdf) of $\Sigma$ for 
$\epsilon=0.5$ (dashed line), $\epsilon=2$ (dotted line), $\epsilon=4$
(dashed-dotted line) and $\epsilon=6$  (full  line).  Upper panel  is
for the whole simulation, middle  panel for cold   gas only and  lower
panel for thermally unstable gas. }
\label{pdf_sig}
\end{figure}

In   the model presented  previously  turbulence  is able to stabilise
transiently a  piece  of thermally   unstable  gas making it able   to
survive longer.  This is done  through the mechanisms presented in the
previous section, which are both related to  the presence of straining
motions  (i.e.  $\Sigma$).  Therefore    this model predicts that  the
thermally  unstable  gas  should   be correlated  with   the parameter
$\Sigma$ in the numerical simulations.  In order to verify this effect
we have smoothed out  the  velocity field at a   scale of 0.2  pc  (10
pixels)  and  computed  $\Sigma$  at every   pixel (Note   that larger
smoothing scales, e.g. 0.4 or 1 pc lead  to similar results).  Then the
correlation between  $\Sigma$    and the fluid   temperature  has been
investigated both globally and locally.

Fig.~\ref{corre_sig_inst}  displays  the total  fraction  of  cold gas
(full line) and the total fraction of cold plus thermally unstable gas
(dashed line)   in the 4  simulations $\epsilon=0.5$,  2, 4  and  6 at
different times as  a function of $<\Sigma  >$.   It is seen that  the
second one is   almost  independent of  $<\Sigma>$ whereas  the  first
decreases linearly  with it.  This result  suggests that  on one hand,
the fraction  of warm gas  which  is driven into the  unstable  regime
depends mainly on  the strength of  the converging flow and is roughly
independent of  the  level of  turbulence.   On  the  other  hand, the
thermally unstable gas  is  able   to  live longer when   $\Sigma$  is
stronger.    For $\epsilon=0.5$, one   has   $<\Sigma> \simeq  1.5  \,
10^{-6}$  year$^{-1}$,  $m  _u/ m_c  \simeq  0.32$   whereas  the mean
pressure of the warm gas before it starts contracting  in CNM is about
3  10$^{-12}$ erg/cm$^3$ (obtained from  Fig.~\ref{press_dens_noturb})
corresponding to   a  mach  number    $M$,  of   $\simeq 1.8$.     For
$\epsilon=4$, $<\Sigma> \simeq   3.5 \,   10^{-6}$, $M  \simeq    1.2$
(obtained  from  Fig.~\ref{press_dens_turb}) and $m  _u  / m _c \simeq
2.8$.  The  relation between the  pressure and   the  Mach number  are
obtained using  isothermal  Rankine-Hugoniot relations,  which seems a
reasonable assumption   looking   at  Fig.~\ref{press_dens_noturb}  and
\ref{press_dens_turb}. The ratio  between $m _u /  m  _c$ for these  2
values is therefore about 8.

Let  us make  a quantitative  comparison  with the  prediction of  the
semi-analytical  model.    Fig.~\ref{corre_sig_inst} predicts for  the
case of  the   simulation  with $\epsilon=0.5$   ($M  \simeq  1.8$ and
$<\Sigma> \simeq 1.5 \, 10^{-6}$  year$^{-1}$),  $\tau _u \simeq  1.5$
Myr  whereas for $\epsilon=4$ ($M   \simeq 1.2$ and $<\Sigma> \simeq
3.5 \, 10^{-6}$ year$^{-1}$) it predicts about 10 Myr.  According to
the arguments presented in Sect.\ref{predict_semana}, this leads to an
estimate for the ratio  between $m_u / m_c$  in these 2 cases of about
10 /  1.5$\simeq$ 7 which  is  in very  reasonable agreement  with the
value ($\simeq 8$) measured in the numerical experiment.

Fig.~\ref{pdf_sig} displays   the pdf  of $\Sigma$ for  $\epsilon=0.5$
(dotted line) and $\epsilon=4$  (full line).  The  upper panel is  for
the whole simulation whereas the middle one  is restricted to cold gas
(T $<$ 200 K) and the lower to thermally unstable gas (200 K $<$ T $<$
5000 K). It is obvious that in the  cold phase, $\Sigma$ is much lower
than for  the  thermally unstable gas  which  is associated  with high
values of $\Sigma$.   Note  that  for $\epsilon=0.5$,  the   thermally
unstable gas has a  lower  $\Sigma$ than  for $\epsilon=4$.  This   is
consistent with the fact that this is turbulence which generates flows
having large value of $\Sigma$.

Both results are, qualitatively and  quantitatively, in good agreement
with  the explanation   proposed in  the   previous sections.  Another
important, though qualitative, fact is that the thermally unstable gas
is  organised  in filaments which   is also a   natural outcome of the
mechanism based on straining motions.

\section{Conclusion}
We perform 2-d numerical simulations with the aim of understanding the
dynamics  of the thermally bistable  interstellar atomic hydrogen. The
size of our  1000$^2$ simulation box is 20  pc, leading to a numerical
resolution of 0.02 pc. Such resolution is  needed in order to properly
describe the cold HI   structures.  In order  to  mimic a large  scale
converging  flow that triggers the transition  of  the warm phase into
the cold one, warm gas is injected  continuously from 2 opposite sides
of  the  box. The gas  can  freely escape at  the  2 other boundaries.
Various amplitudes of turbulent fluctuations have been applied.

When the flow is weakly turbulent, a layer of compressed WNM forms and
quickly fragments into structures of cold gas.   When the flow is very
turbulent,  it is less organised.  Nevertheless the thermally bistable
behaviour  is not  erased  and  indeed  locally  very  similar to  the
classical picture of the 2-phase medium.   In particular, the 2 phases
are connected  through sharp thermal fronts and   are locally in rough
pressure equilibrium.

In order to  characterise the CNM  structures  found in  the numerical
simulation, we apply a simple threshold algorithm to identify them and
give  the pdf of  some of their  intrinsic  properties as mean density,
temperature, velocity dispersion,  size   and aspect ratio, $r$.    Typical
values for  these parameters  are   respectively: $<n> \simeq   50$
cm$^{-3}$, T$\simeq$80 K, $\delta v \simeq$ 0.5 km/s, l$\simeq$ 0.1 pc
and r$\simeq$0.5.

A large fraction  of thermally unstable gas  which increases with  the
amplitude of the turbulent forcing,  is found. This thermally unstable
gas tends to be organised in filamentary structures.

A semi-analytical model for  the thermal and  dynamical evolution of a
fluid particle is presented. It is based on the Lagrangian description
of a  fluid element.  This  model predicts  that substantial straining
motions may  efficiently stabilise a piece   of thermally unstable gas
allowing it to survive a  longer period of time.   We then verify than
both locally and globally the  thermally unstable gas in the numerical
simulations  is  indeed  very  strongly  correlated  with  substantial
straining motions.

\section{acknowledgements}
We acknowledge  the support of  the CEA computing center,  CCRT, where
all the  simulations where  carried out. We  also would like  to thank
Jean-Michel   Alimi   and   Jean-Pierre   Chi\`eze   for   stimulating
discussions.  We  acknowledge a critical reading of  the manuscript by
Michel  P\'erault as  well as  enlighting discussions  on  the results
presented in this manuscript.   PH greatly acknowledges the support of
a CNES fellowship.

\appendix

\section{A semi-analytical model}
In this  appendix we derive the  semi-analytical model that  we use in
Sect.~\ref{turb_therm}. The  idea of  the model is  to follow  a fluid
element by  writing Eqs.~\ref{mcons}-\ref{econs} in  a Lagrangian form
and  then,  making  some   approximations,  to  compute  the  pressure
gradients.

The conservation of the mass can be written in a Lagrangian form as:
\begin{equation}
\label{equ_dens}
\dot{\rho} + \rho \theta = 0.
\end{equation}
where $\theta$ is the  divergence of the  velocity  field and the  dot
represent  a total   (i.e.  Lagrangian) derivative.  The evolution  of
thermal energy can also be simply written in a Lagrangian form:
\begin{equation}
\label{equ_ener}
\dot{\cal{E}} + \frac{P}{\rho} \theta =  \frac{1}{\rho} {\cal L}(\rho,T),
\end{equation}
where ${\cal{E}} = P/(\rho(\gamma-1))$ is the specific thermal energy.

In order  to have  an  evolution  equation  for $\theta$ we  take  the
gradient of Eq.~(\ref{momcons}). As it is common in fluid mechanics,
we  then decompose  the resulting tensorial   equation into its trace,
symmetric trace-free  and    anti-symmetric  part. For   the  sake  of
simplicity and since  we want to compare  our model to  the simulation
presented above, we write these equations for a bi-dimensional flow :

\begin{eqnarray}
\dot{\theta}   + & \frac{\theta^2}{2} + 2(\sigma_1^2 + \sigma_2^2 -\omega^2)& = - (P_{xx}+P_{yy}), \\
\label{equ_s1}
\dot{\sigma_1} + & \theta \sigma_1 & = -\frac{1}{2} (P_{xx} - P_{yy}),  \\
\dot{\sigma_2} + & \theta \sigma_2 & = -\frac{1}{2}(P_{xy} + P_{yx}), \\
\label{equ_rot}
\dot{\omega}   + & \theta\omega + \omega(\sigma_2 - \sigma_1) & = 0,
\end{eqnarray}
where $\sigma_1 = 0.5(\partial_x  u_x - \partial_y u_y)$ and $\sigma_2
= 0.5(\partial_x u_y + \partial_y u_x)$  are the two components of the
shear tensor and  $\omega = 0.5(\partial_x  u_y - \partial_y u_x)$  is
the rotational.    The   $P_{ij}$   are  defined   by :     $P_{ij}  =
\partial_i(\partial_j P/\rho)$. 

The system (\ref{equ_dens})-(\ref{equ_rot}) is exact but is not closed
since there is no evolution equation for $P_{ij}$. Therefore it cannot
be  integrated  unless  one makes  some  approximation  to compute the
$P_{ij}$.

The simplest case  to consider, as a starting  point, is the evolution
of  an isobaric  fluctuation.   In that case  $P_{ij}  =   0$ and  the
evolution of the fluid element is given by:

\begin{equation}
\theta = -\frac{\dot{\rho}}{\rho} = \frac{(\gamma-1)}{\gamma P_0}  {\cal L}(\rho,T) 
= -\frac{\dot{\sigma_1}}{\sigma_1}= -\frac{\dot{\sigma_2}}{\sigma_2}
\end{equation}

where $P_0$ is the constant pressure and $e_0 = P_0/(\gamma-1)$ is the
corresponding   internal energy.    The cooling   time, $\tau_{cool} =
-e_0/{\cal L}(\rho,T)$ is the characteristic time  of evolution of the
system.

In order to   treat the pressure gradient  let  us first simplify  the
previous  system.  If rotation is  neglected, the shear  tensor can be
diagonalized  and  the eigenvector basis  will   not rotate during the
evolution. In this eigenframe, the only  component of the shear tensor
is given by $\Sigma = \sqrt{\sigma _1^2  + \sigma _2^2}$ and the fluid
element can  be viewed as an ellipsoid  defined by its  semi-major and
minor  axis  of length $a$  and  $b$.  The  equation  of evolution for
$\Sigma$ is identical to that  of  $\sigma_1$ (\ref{equ_s1}) with  the
pressure  gradient  taken in  the  eigenframe.  The  evolution of  the
ellipsoid axis is given by:

\begin{equation}
a(t) = a_0 \exp \left( \int_0^t (\theta/2 + \Sigma) dt \right)  \equiv a_0
a_{\theta} a_{\Sigma} \; , \; 
b(t) = b_0 \exp \left( \int_0^t (\theta/2 - \Sigma) dt \right)
  \equiv b_0 a_{\theta}/a_{\Sigma}
\label{evo_axe}
\end{equation}
where   we have   defined  $a_{\theta}  =  \exp( \int_0^t \theta/2 dt)
$  and
$a_{\Sigma} = \exp( \int_0^t \Sigma  dt) $.

If we further assume that the pressure  and density inside and outside
the ellipsoid are uniform then we may write:
\begin{equation}
P_{xx} \simeq \frac{ P_{\rm ext} - P(t) }{\rho(t) a(t)^2}, \qquad   P_{yy} \simeq
\frac{  P _{\rm ext} - P(t)  }{\rho(t) b(t)^2} \;\mbox{ and }\; P_{xy} \simeq  0.
\end{equation}

Using this approximation for $P_{ij}$, it is now possible to integrate
the system (\ref{equ_dens})-(\ref{equ_rot}) to determine the evolution
of the fluid element.

  

\end{document}